\documentclass[useAMS,usenatbib]{mn2e}
\usepackage{natbib,hyperref}
\usepackage{epsfig,graphicx}
\usepackage{color}
\newcommand{\hto}{\widehat{t}_o}

\newcommand{\x}{{\bf x}}
\newcommand{\X}{{\bf X}}
\newcommand{\y}{{\bf y}}

\newcommand{\hf}{\widehat{f}_{\rm Ia}}
\newcommand{\hp}{\widehat{p}_{\rm Ia}}
\newcommand{\he}{\widehat{e}_{\rm Ia}}
\newcommand{\tstar}{t_{\rm Ia}^*}
\newcommand{\tclass}{\gamma_{\rm Ia}}
\newcommand{\estar}{\epsilon^*}
\renewcommand{\P}{{\bf P}}

\newcommand{\new}[1] {\textcolor{black} {{#1}}}

\title[Semi-supervised SN Classification]{Semi-supervised Learning for Photometric Supernova Classification\thanks{A web service for the supernova classification method used in this paper can be found at \url{http://supernovaclass.info/}}}
\author[J. W. Richards et al.]
       {\parbox[]{6.0in}
{Joseph W.\ Richards$^{1,2}$\thanks{E-mail:
jwrichar@stat.berkeley.edu (JWR)}, Darren Homrighausen$^{3}$, Peter E.\ Freeman$^{3}$,
Chad M.\ Schafer$^{3}$, and Dovi Poznanski$^{1,4}$\\
        \footnotesize
$^{1}$Department of Astronomy, University of California, Berkeley, CA, 94720-7450, USA\\
$^{2}$Department of Statistics, University of California, Berkeley, CA, 94720-7450, USA\\
$^{3}$Department of Statistics, Carnegie Mellon University, 5000 Forbes Avenue, Pittsburgh, PA 15213, USA\\
$^{4}$Computational Cosmology Center, Lawrence Berkeley National Laboratory, 1 Cyclotron Road, Berkeley, CA 94720, USA}}

\begin{document}

\date{Accepted 2011 September 5. Received 2011 March 28 ; in original form 2011 March 28}

\pagerange{\pageref{firstpage}--\pageref{lastpage}} \pubyear{2011}

\maketitle
\label{firstpage}

\begin{abstract}
We present a semi-supervised method for photometric supernova typing.  Our approach is to first use the nonlinear dimension reduction technique diffusion map to detect structure in a database of supernova light curves and subsequently employ random forest classification on a spectroscopically confirmed training set to learn a model that can predict the type of each newly observed supernova.  We demonstrate that this is an effective method for supernova typing. As supernova numbers increase, our semi-supervised method efficiently utilizes this information to improve classification, a property not enjoyed by template based methods.  Applied to supernova data simulated by \citet{kess2010} to mimic those of the Dark Energy Survey, our methods achieve  (cross-validated) 95\% Type Ia purity and 87\% Type Ia efficiency on the spectroscopic sample, but only 50\% Type Ia purity and 50\% efficiency on the photometric sample due to their spectroscopic follow-up strategy.  To improve the performance on the photometric sample, we search for better spectroscopic follow-up procedures by studying the sensitivity of our machine learned supernova classification on the specific strategy used to obtain training sets.  With a fixed amount of spectroscopic follow-up time, we find that, despite collecting data on a smaller number of supernovae, deeper magnitude-limited spectroscopic surveys are better for producing training sets.  For supernova Ia (II-P) typing, we obtain a 44\% (1\%) increase in purity to 72\% (87\%) and 30\% (162\%) increase in efficiency to 65\% (84\%) of the sample using a 25th (24.5th) magnitude-limited survey instead of the shallower spectroscopic sample used in the original simulations.  When redshift information is available, we incorporate it into our analysis using a novel method of altering the diffusion map representation of the supernovae. Incorporating host redshifts leads to a 5\% improvement in Type Ia purity and 13\% improvement in Type Ia efficiency.
\end{abstract}


\section{Introduction}
\label{sec:intro}

Novel approaches to photometric supernova (SN) classification are in high demand in the astronomical community.  The next generation of survey telescopes, such as the Dark Energy Survey (DES; \citealt{des}) 
and the Large Synoptic Survey Telescope (LSST; \citealt{ivez2008}), are expected to observe light curves for a few hundred thousand supernovae (SNe), far surpassing the resources available to spectroscopically confirm the type of each. To fully exploit these large samples, it is necessary to develop methods that can accurately and automatically classify large samples of SNe based only on their photometric light curves.


In order to use Type Ia supernovae as cosmological probes, it is imperative that pure and efficient Type Ia samples are constructed.  Yet, classifying SNe from their light curves is a challenging problem.  The light flux measurements are often noisy, nonuniform in time, and incomplete.  In particular, it is difficult to discern the light curves of Type Ia SNe from those of Type Ib or Ic supernovae, explosive events which result from the core collapse of massive stars.  This difficulty can have dire effects on the subsequent cosmological inferences:  if the sample of SNe Ia used to make cosmological inferences is impure, then the cosmological parameter estimates can be significantly biased (\citealt{home2005}).  On the other hand, if the classification of Type Ia SNe is not efficient, then cosmologists do not fully utilize the sample of supernovae on hand, resulting in a loss of precision.

In the last decade, several supernova photometric classification methods have been introduced.  These include the methods of \citet{pozn2002,sull2006,john2006,pozn2007,kuzn2007,kunz2007,rodn2009,gong2010}, and \citet{falc2010}.  Each of these approaches uses some version of template fitting, where each observed supernova's data is compared to the data from well established SNe of different types and the subsequent classification is estimated as the SN type whose template fits best (usually judged by maximum likelihood or maximum posterior probability).  Usually, the sets of templates are constructed using the observed spectra of sets of well studied, high signal-to-noise SNe (see \citealt{nuge2002}) or spectral energy distribution models of SNe. 

The principal drawback of using template methods for classification is that they depend heavily on the templates used for each of the different classes.  If there are errors in the template creation, these will propagate down to the estimated classifications.  Furthermore, template fitting often assumes that each observed SN can be well modeled by one of the templates, an assumption that may be unreasonable, especially for large data sets.  Additionally, these methods require that all relevant parameters (such as redshift and extinction) in the light curve model be simultaneously fit or otherwise estimated, which can pose computational challenges and cause catastrophic errors when the estimates are poor.  

A viable alternative to template fitting for SN classification is {\it semi-supervised learning},
a class of methods that are able to learn the low dimensional structure
in all  available data and exploit that structure in classification; as more data are obtained,
these methods are able to utilize that additional information to better classify
{\bf all} of the SNe.
Template fitting methods, on the other hand, do not automatically learn as
new data are collected 
\new{(for example, \citealt{2011arXiv1107.5106S} extract only 8 core-collapse SN templates from over 10,000 observed supernova candidates).}
Adverse effects caused by incorrectly classified templates, 
under-sampled regions of template parameter space, or unrealistic templates
will not go away as
the number of SNe increases.  Indeed, the magnitude of these biases relative
to the statistical errors in the estimates will only increase. Well constructed
semi-supervised approaches will reduce both bias and variance as the
sample sizes grow.

Our main contribution in this paper is to introduce a method for SN photometric classification that does not rely on template fitting.  Our proposed method uses semi-supervised learning on a database of SNe: we first use all of the light curves to simultaneously estimate an appropriate, low dimensional representation of each SN, and then we employ a set of labeled (spectroscopically confirmed) examples to build a classification model in this reduced space, which we subsequently use to estimate the type of each unknown SN. 

An advantage to our semi-supervised approach is that it learns from the set of unlabeled SNe.  Typically there are many more unlabeled than labeled supernovae, meaning that a method that learns from all the data is an obvious improvement over methods that only learn from the labeled SNe.  Another advantage is that our method gives an accurate prediction of the class of each supernova without having to simultaneously estimate nuisance parameters such as redshift, stretch or reddening.  This result arises because variations in these parameters appear as gradual variations in the low dimensional representation of the light curves when the observed data (labeled plus unlabeled SNe) are collected densely enough over the nuisance parameters.  In the low dimensional representation, the variability due to supernova type is greater than the variability due to the nuisance parameters, allowing us to build an accurate classifier in this space.

Until recently, there had been no standard testing procedure for the various SN classification methods. To assess the performance of these methods, \citet{kess2010} held a public ``SN Photometric Classification Challenge" in which they released a blended mix of simulated supernovae (Ia, Ib, Ic, II) to be classified.  As part of the Challenge, a spectroscopically confirmed subset of SNe was included on which the Challenge participants could train or tune their algorithms.  The results of that Challenge (\citealt{kess2010a}) showed that various machine learning classification algorithms were competitive with classical template fitting classifiers.   Apart from our entry (InCA\footnote{International Computational Astrostatistics Group, \url{http://www.incagroup.org}}), none of the other entries to the Classification Challenge used semi-supervised learning.   In this paper we will further explore the viability of semi-supervised classifiers for SN classification.

Recently, \citet{newl2010} released a paper detailing their entries into the SN Challenge.  They argue that their methods are comparable to the best template fitting techniques, but that they require training samples that are representative of the full (photometric) sample.  Our findings are similar, and we carry out a detailed sensitivity analysis to determine how the accuracy of predicted SN types depend on characteristics of the training set.  Like \citet{newl2010}, we  perform our analysis both with and without photometric redshift estimates; we introduce a novel and effective way of using photo-z estimates in finding a low dimensional embedding of SN data.


Based on our sensitivity analysis, we conclude that magnitude-limited spectroscopic follow-up strategies with deep limits (25th mag) produce the best training sets for our supernova classification method, at no extra observing cost.  Though these deeper observing strategies result in fewer supernovae than shallower samples, they produce training sets that more closely resemble the entire population of SNe under study, causing overwhelming improvements in supernova classification.  We strongly recommend that spectroscopic SN follow-up is performed with faint magnitude limits.

The layout of the paper is the following. In {\S}\ref{sec:meth}, we describe our semi-supervised classification method, which
is based on the diffusion map method for nonlinear dimensionality reduction
and the random forest classification technique.  We couch our description
in general terms since elements of our methodology can in principle be applied
to any set of light curves, not just those from supernovae.  
In {\S\S}\ref{sec:snapp}-\ref{sec:snapp1}, we describe the application of our methodology
to the problem of classifying supernovae, fully detailing the steps in our analysis.  These sections
are divided into the unsupervised (\S\ref{sec:snapp}) and supervised (\S\ref{sec:snapp1}) parts of the analysis.
We assess the performance of our methods in \S\ref{sec:results}, using the 21,319 light curves
simulated for the SN Photometric Classification Challenge (\citealt{kess2010}).
We examine the sensitivity of our results to the composition of
the training datasets, and to the treatment of host-galaxy redshift (by either using redshifts to alter the diffusion map construction or
using them as covariates in random forest classification).  In {\S}\ref{sec:summary}, we
offer a summary and our conclusions.
We provide
further information on classification trees and the random forest
algorithm in Appendices \ref{ss:classificationTrees} and \ref{ss:randomForests}.

\section{Methodology: Semi-Supervised Classification}
\label{sec:meth}

Suppose we have observed photometric light curves for $N$ objects.
For each, the observed data are flux measurements at irregularly-spaced 
time points in each of a number of different filters (e.g., {\it griz}), 
plus associated measurement errors.  Call the data for object $i$, 
$\X_i = \{t_{ik}^b, F^b_{ik},\sigma_{ik}^b\}$, where $k=1,...,p_i^b$, 
$b$ indexes the filter, and $p_i^b$ is the number of flux measurements 
in filter $b$ for object $i$.  Here, $\mathbf{t}_{i}^b$ is the time grid, and
$\mathbf{F}_i^b$ and $\mathbf{\sigma}_i^b$ are the flux measurements and errors, respectively, in filter $b$.
Suppose, without loss of generality, that the first $n$ objects, 
$\X_1,...,\X_n$, have known types $y_1,...,y_n$.
Our goal is to predict the type of each of the remaining $N-n$ objects.

To perform this classification, we take a \emph{semi-supervised} approach 
(see \citealt{chap2006} for an introduction to semi-supervised learning 
methods).  The basic idea is to use the data from all $N$ objects (both labeled and 
unlabeled) to find a low dimensional representation of the data 
(the unsupervised part) and then to use just the $n$ labeled objects to train a 
classifier (the supervised part).  Our proposed procedure is as follows:
\begin{enumerate}
\item Do relevant preprocessing to the data.  Because this is 
application specific, we defer details on our preprocessing of SN light curves
to the next section.  Here, it suffices to state that the result of this
preprocessing for one object is 
a set of light curve function estimates interpolated over a fine time grid.
\item Use the preprocessed light curves for all $N$ objects to learn 
a low dimensional, data driven embedding of $\{\X_1,...,\X_N\}$.  We 
use the diffusion map method for nonlinear dimensionality reduction
(\S\ref{ss:dmap}).
\item Train a classification model on the $n$ labeled (spectroscopically confirmed) examples that 
predicts class as a function of the diffusion map coordinates 
of each object.  We use the random forest method (\S\ref{ss:rf}) as 
our classifier.
\item Use the estimated classification model to predict the type of 
each of the $N-n$ unlabeled objects.
\end{enumerate}
The advantage of the semi-supervised approach is that it uses all of 
the observed data to estimate the lower dimensional structure of the object
set.  Generally, we have many more objects without classification labels 
than with (e.g, the SN Photometric Classification Challenge provided 
$14$ unlabeled SNe per labeled SN).  If we have continuously varying 
parameters that affect the shapes and colours of the light curves
(e.g., redshift, extinction, etc.) then it is imperative that we 
use as many data points as possible to capture these variations when 
learning a low dimensional representation.  We then use the examples 
for which we know the object type to estimate the classification model, 
which is finally employed to predict the type of each unlabeled object. 

\subsection{Diffusion Map}
\label{ss:dmap}

In this section, we review the basics of the diffusion map approach to spectral
connectivity analysis (SCA) for 
data parametrization.  We detail our approach of using diffusion map to 
uncover structure in databases of astronomical objects. 
For more specifics on the diffusion map technique, we refer the reader to 
\citet{coif2006} and \citet{lafo2006}.  For examples of the application of 
diffusion map to problems in astrophysics, see \citet{rich2009a}, 
\citet{rich2009b}, and \citet{free2009}.  In \citet{rich2009a}, the authors compare and contrast the use of diffusion maps, which are non-linear, with a more commonly utilized linear technique, principal components analysis (PCA), and demonstrate the superiority of diffusion maps in predicting spectroscopic redshifts of SDSS data.

The basic idea of diffusion map is the following. To make statistical prediction (e.g., of SN type) tractable, one seeks a simpler parameterization of the observed data, which is often complicated and high-dimensional. The most common method for data parameterization is PCA, where the data are projected onto a lower-dimensional hyperplane. For complex situations, however, the assumption of linearity may lead to suboptimal predictions because a linear model pays very little attention to the natural geometry and variations of the system.

The top plot in Figure 1 of \citet{rich2009a} illustrates this  by showing a data set that forms a one-dimensional noisy spiral in two dimensional Euclidean space. Ideally, we would like to find a coordinate system that reflects variations along the spiral direction, which is indicated by the dashed line. It is obvious that any projection of the data onto a line would be unsatisfactory.  If, instead, one imagines a random walk starting at $\x$ that only steps to immediately adjacent points, it is clear that the number of steps it takes for that walk to reach $\y$ reflects the length between $\x$ and $\y$ along the spiral direction. 
This is the driving idea behind the diffusion map, in which the ``connectivity" of the data in the context of a fictive diffusion process, is retained in a low-dimensional parametrization.  This simple, non-linear parametrization of the data is useful for uncovering simple relations with the quantity of interest (e.g., supernova type) \new{and is robust to random noise in the data}.

We make this more concrete below.


Diffusion map begins by creating a weighted, undirected graph on our 
observed photometric data $\{\X_1,...,\X_N\}$, where each data point 
is a node in the graph and the pairwise weights between nodes are defined as
\begin{equation}
\label{eqn:dmap1}
w(\X_i,\X_j) = \exp\left(-\frac{s(\X_i,\X_j)}{\epsilon}\right)
\end{equation}
where $\epsilon >0$ is a tuning parameter and $s(\cdot,\cdot)$ is a user-defined
pairwise distance between objects.  Here, $s$ is a
\emph{local} distance measure, meaning that it should be small only if $\X_i$ and $\X_j$
are similar  (in \S\ref{ss:lcdist} we
define the distance measure we use for SN light curve data).  In this construction, 
the probability of stepping from $\X_i$ to $\X_j$ in one step of a diffusion process
is $p_1(\X_i, \X_j) = w(\X_i,\X_j)/ \sum_k w(\X_i, \X_k)$.
We store the one step probabilities between all $N$ data points 
in the $N \times N$ matrix $\P$; then, by the theory of Markov chains, 
for any positive integer $t$, the element $p_t(\X_i, \X_j)$ of the matrix
power $\P^t$ gives the probability of going from $\X_i$ to $\X_j$ in $t$ steps.
See, e.g., Chapter 6 in \citet{grim2001} for an introduction to Markov chains.

We define the diffusion map at scale $t$ as
\begin{equation}
\label{eqn:dmap2}
\mathbf{\Psi}^t : \X \mapsto \left[ \lambda_1^t \mathbf{\Psi}_1(\X), \lambda_2^t \mathbf{\Psi}_2(\X),..., \lambda_m^t \mathbf{\Psi}_m(\X)\right]
\end{equation}
where $\mathbf{\Psi}_j$ and $\lambda_j$ are the right eigenvectors 
and eigenvalues of $\P$, respectively, in a biorthogonal spectral decomposition
and $m$ is the number of diffusion map coordinates chosen to represent the data.  
\new{The diffusion map coordinates are ordered such that $\lambda_1 \ge \lambda_2 \ge ... \ge \lambda_m$, so
that the top $m$ coordinates retain the most information about $\mathbf{P}^t$.  Though there are $N$ total
eigenvectors of $\mathbf{P}$, only $m \ll N$ are required to capture most of the variability of the system.}

The Euclidean distance between any two points in the $m$-dimensional space 
described by equation (\ref{eqn:dmap2}) approximates the diffusion distance, 
a distance measure that captures the intrinsic geometry of the data set by 
simultaneously considering all possible paths between any two data points 
in the $t$-step Markov random walk constructed above.  
\new{Because it averages over all possible paths between data points in the random walk,
the diffusion distance is robust to noise in the observed data.}
The choice of the parameters
$\epsilon$ (in equation \ref{eqn:dmap1}) and $m$ gives the map a great deal of flexibility, and it is
feasible to vary these parameters in an effort to obtain the best classifier
via cross-validation, as described in \S\ref{ss:classtune}.

We note that in the random forest classifier used to predict object type 
(see \S\ref{ss:rf}), the scale of each coordinate of $\mathbf{\Psi}^t(\X)$ 
does not influence the method because each classification tree is constructed 
by splitting one coordinate at a time.  Therefore, the parameter $t$, 
whose role in the diffusion map (\ref{eqn:dmap2}) is to rescale each 
coordinate, has no influence on our analyses.  We choose to fix $t$ to 
have the value 1.  For the remainder of this paper, we will use 
$\mathbf{\Psi}_i$ to stand for the $m$-dimensional vector of diffusion 
map coordinates, $\mathbf{\Psi}^1(\X_i)$.

\subsection{Classification: Random Forest}
\label{ss:rf}

In finding the diffusion map representation of each object, the idea
is that this parameterization will hopefully obey a simple relationship with respect to object
type.  Then, simple modeling can be performed to build a class-predictive
model from the diffusion map coordinates.
That is, once we have the $m$-dimensional diffusion map representation of each
object's light curve (equation \ref{eqn:dmap2}), we construct a model
to predict the type, $y_i$, of the $i^{th}$ object as a function of its
diffusion map representation, $\mathbf{\Psi}_i$.  In other words,
using the set of $n$ SNe with known classification labels, we estimate
the underlying function $h$ that relates each $m$-dimensional
diffusion map representation with a classification label.  We will
ultimately use this estimate, $\widehat{h}$, to predict the
classification, $\widehat{y}_j = \widehat{h}(\mathbf{\Psi}_j)$ for
each unlabeled supernova $j=n+1,...,N$.

Any classification procedure that can handle more than
two classes could be applied to the diffusion map representation of the
SNe. By adapting the predictors to the underlying structure in the data,
standard classification procedures should be able to separate the SNe of
different types.  
\new{There are many classification tools available in the statistics and machine
learning literature, ranging to simple $K$ nearest-neighbor averaging to more
sophisticated ensemble methods (see \citealt{2011arXiv1104.3142B} for an 
overview of some of the classification methods that have been used in time-domain astronomy).}
We choose to use the \emph{random forest} method of 
\citet{brei2001} due to its observed
success in many multiclass classification settings, including in astrophysics
(\citealt{2003sca..book..243B,rich2011,2011arXiv1101.2406D}).  
The basic idea of the random forest is to build a large collection of 
decorrelated classification tree estimates and then to average these estimates to obtain the final predictor, 
$\widehat{h}$.  
This approach usually works well because classification tree estimates 
are notoriously noisy, but tend to be unbiased.  By averaging decorrelated 
tree estimates, the random forest produces classification estimates 
that are both unbiased and have small variance with respect to the 
choice of training set.  
See the Appendices A and B
for a brief overview of classification trees 
(\S\ref{ss:classificationTrees}) and random forest (\S\ref{ss:randomForests}).

\section{Semi-Supervised Supernova Classification: The Unsupervised Part}
\label{sec:snapp}

The first step of our analysis is to use the entire dataset of supernova light
curves to learn an appropriate representation of each supernova using diffusion map.  This is the
\emph{unsupervised} part of our semi-supervised approach.

\subsection{Data}
\label{ss:data}

We apply the methods described in {\S}\ref{sec:meth} to 
the {\tt SNPhotCC+HOSTZ} dataset of the
SN Photometric Classification Challenge (\citealt{kess2010}).  We use data from the ``Updated Simulations"\footnote{Data can be downloaded at \url{http://sdssdp62.fnal.gov/sdsssn/SIMGEN_PUBLIC/SIMGEN_PUBLIC_DES.tar.gz}}, as described
in \S6 of \citet{kess2010a}.  These data were simulated to mimic the observing conditions of the DES.  Note that these data are significantly different than the data used in the Challenge, with several bug fixes and improvements to make the simulations more realistic.   For instance, the ratio of photometric to spectroscopic SNe was 13:1 in the Challenge data, but 18:1 in the Updated Simulations.  Therefore, we refrain from comparing our current results directly to the results in the Challenge.\footnote{Specifically, we find that our methods perform 40\% better on the Challenge data than on the data used in this paper in terms of the photometric Ia figure of merit.} 

We denote the updated Challenge data as $\mathcal{D}$.  There are a total of 
$N$ = 20,895 SNe in $\mathcal{D}$\footnote{After removal of the 424 SNe simulated from the SDSS 2007nr II-P template, whose 
peak luminosities are anomalously dim by several magnitudes.} and for each SN we have 
{\it griz} photometric light curves.  These light curves are simulated 
by the Challenge 
organizers so as to mimic the observing conditions of the Dark 
Energy Survey (DES; \citealt{bern2009}).  The light curve for each SN is measured on anywhere between
16 to 160 epochs (between 4 and 40 epochs in each filter), with a median value of 101 epochs.
The maximum $r$-band signal-to-noise ratio for the light curves ranges between 0.75 and 56, with a median of 11.

The data, $\mathcal{D}$, is originally comprised of two sets.  One that we dub $\mathcal{S}$
contains 1,103 spectroscopically confirmed SN light curves, while the 
other, containing 19,792 photometrically observed SNe, is dubbed $\mathcal{P}$.

%

\subsection{Data preprocessing}
\label{ss:preprocessing}

In order to utilize diffusion map in our classification scheme, we
first need to formulate a distance metric, $s$, between observed SN light curves
(see equation \ref{eqn:dmap1}).  Our distance measure
should capture differences in the shapes of the light curves and the colours of the SN.  Phrased differently,
because the measure $s$ is only considered on small scales (controlled by $\epsilon$
in equation \ref{eqn:dmap1}), we wish to construct a measure that is small only if
\emph{both} the shapes of the light curves and the colours of two SNe are very similar.


\subsubsection{Non-Parametric Light Curve Shape Estimation}

Each SN is sampled on an irregular time grid that differs from 
filter to filter and from SN to SN.  To circumvent this difficulty,
we find a nonparametric estimate of the shape of the 
observed light curve, $\X_i = \{t_{ik}^b, F^b_{ik},\sigma_{ik}^b\}$, for each SN.  With this estimate,
we can shift from using the irregularly sampled observed data to
using fluxes on a uniform time grid when computing the
distances between light curves.

We independently fit a natural cubic regression spline to the 
data from each filter, $b$, and each SN, $i$ (see, e.g., 
\citealt{rupp2003} and \citealt{wass2006}).
We utilize
regression splines because they are particularly useful for estimating smooth,
continuous functions that can have complicated behavior such as 
rapid increases.  In doing so, we avoid assuming an overly restrictive template
model for the shape of each light curve.  We also leave the number of spline knots
as a free parameter, allowing the model to adapt to the complexity of the true
SN light curve shape.   Our cubic spline estimator is
\begin{equation}
\label{eqn:spline}
\widehat{F}_{ik}^b \equiv \widehat{F}_i^b(\widehat t_{k}) 
=  \sum_{j=1}^{\nu + 4} \widehat{\beta}_{ij}^b B_j(\widehat t_{k})
\end{equation}
where $B_j$ is the $j^{th}$ natural cubic spline basis, $\widehat t_{k}$ are
points along our uniform time grid, and $\nu$ is the number of knots.
The $\widehat{\beta}_{ij}^b$ are estimators of the coefficients $\beta_{ij}^b$ 
and are fit from the observed $\X_i$ by minimizing weighted least 
squares against the observed fluxes $F_{ik}^b$ with weights $(1/\sigma_{ik}^b)^2$.
By fiat, we choose
a grid of 1 measurement per MJD, noting that we have found denser grids 
to have negligible effect on our final results while incurring increased computing time.
Other implementations could use a sparser grid, resulting in faster computing times.

When fitting regression splines, we must choose
the quantity, $\nu$, and locations of the knots,
which correspond to points of discontinuity of the third derivative of the 
spline function.
We follow convention and place the knots uniformly over the observed time points.  
To choose $\nu$,
we minimize the generalized cross-validation (GCV) score, defined as
\begin{equation}
\label{eqn:gcv}
{\rm GCV}_{i,b}(\nu) = \frac{1}{p_i^b}\sum_{k=1}^{p_i^b}\left(\frac{F_{ik}^b-\widehat{F}_{ik}^b}{\sigma_{ik}^b(1-\nu/p_i^b)}\right)^2
\end{equation}
where $\widehat{F}_{ik}^b$ is the fitted value of the spline with $\nu$ knots at
$\widehat t_{k}$ 
computed using eq.~(\ref{eqn:spline}).
Minimizing eq. (\ref{eqn:gcv}) over 
$\nu$ balances the bias and variance (bumpiness) of the fit.  
Note that the observational uncertainties in the measured fluxes, $\sigma_{ik}^b$, are used  to
compute both the LC flux estimates, $\widehat{F}_{ik}^b$, and the model uncertainty in those estimates, $\widehat{\sigma}_{ik}^b$.  In \S\ref{ss:lcdist}
we will use these entities to construct a distance metric used to compute the diffusion map representation of each SN.

\begin{figure}
\includegraphics[angle=0,width=3.5in]{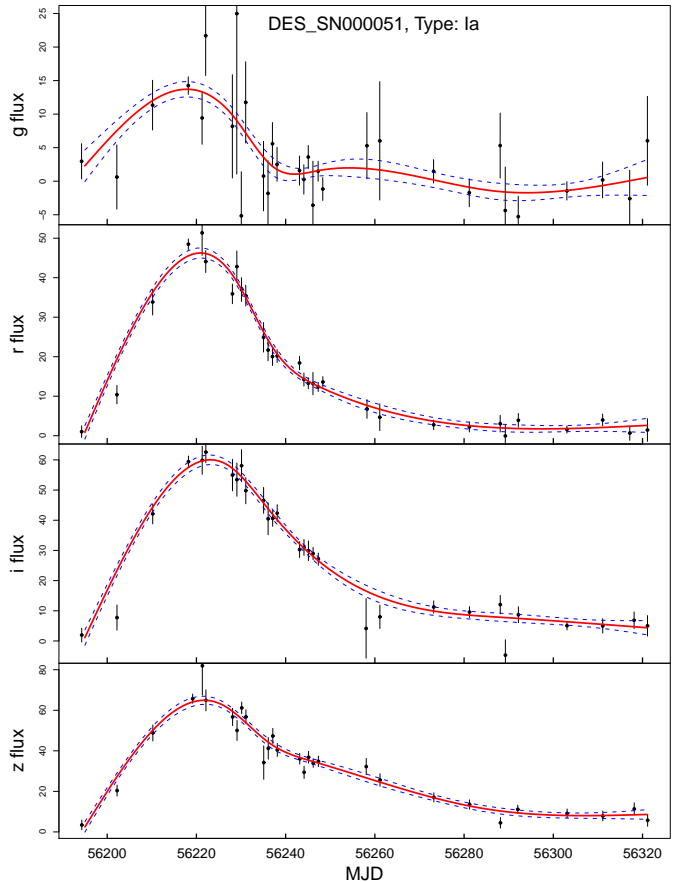}
\caption{Spline fit to the data of a single Type Ia supernova.  A regression spline was fit independently to the light curve data in each band using GCV (equation \ref{eqn:gcv}).  The spline fit (solid red line) and model errors in that fit (blue dashed lines) are used to compute a pairwise supernova light curve distance (\ref{eqn:distband}), which is used to find a diffusion map representation of each SNe.
}
\label{fig:LC}
\end{figure}

We have applied 
the spline fitting routine to a set of simulated light curves, and the method 
produces reasonable fits.   
For an example of the fit for a single 
supernova's {\it griz} light curves, see Figure~\ref{fig:LC}.
For a few SNe, 
minimizing the GCV led to too many knots (the estimated light curve was 
deemed too bumpy), so the parameter $\nu$ is restricted to be no greater 
than 10 to ensure that the estimated curve for each band is physically plausible.

Once we estimate a SN's raw light curve function, we normalize the
flux to mitigate the effect that the observed brightness of each SN has on the 
pairwise distance estimates.  The normalized flux of SN $i$ is
\begin{equation}
\widetilde{F}_{ik}^b \equiv \frac{\widehat{F}_{ik}^b}{\sum_{b \in griz} \max_k \{\widehat{F}_{ik}^b\}} \,.
\end{equation}
Similarly, we normalize the model error,  $\widehat{\sigma}_{ik}^b$, associated with each spline-estimated flux
$\widehat{F}_{ik}^b$; call this $\widetilde{{\sigma}}_{ik}^b$.  Because the same divisor is used for each
photometric band,  the colour of each supernova is preserved. A distance
measure constructed from $\{\widetilde{{F}}_i^b,\widetilde{{\sigma}}_i^b\}$ will capture both the shape and
colour of the light curve.

\subsubsection{Zero-Point Time Estimation}

\new{In order to accurately measure the similarities and differences in the shapes of the observed
light curves, we must ensure that they are aligned on the time axis.
To this end, we define the \emph{zero-point time} of a SN light curve as its time, in MJD, of
maximum observed  {\it r}-band flux, $t_{o,i} = \max_k \{\widehat{F}_{ik}^r\}$.  Shifting all the light curves to this common
frame, but setting $\widetilde{t}_i = \widehat{t}_i - t_{o,i}$ enables us to construct a pair-wise distance measure that captures differences in observed LC shapes and colors.}

\new{We estimate the zero-point time for a SN whose {\it r}-band maximum
occurs at a time endpoint---either the first or last epoch of the observed light 
curve---on a case by case basis: if it is being compared to
a light curve whose maximum {\it r}-band flux occurs either at the same endpoint or at neither endpoint,
we use cross-correlation to estimate $t_o$; if it is
being matched with a SN whose {\it r}-band flux occurs at the opposite endpoint, we
abort the zero-point estimation process and set the distance to a large
reference quantity since the two light curves are, by definition, dissimilar.}
In our SN database, 2908 (14\%) of the supernovae  are observed post $r$-band peak
while 1489 (7\%) are observed pre $r$-band peak.

\new{Our use of {\it r}-band data to establish a zero-point time is motivated
by simulations of Type Ia SNe data.
Let $\widehat{t}_o$ denote our estimator of  $t_o$.
Using 1,000 SNeIa simulated with {\tt SNANA} (\citealt{Kessler09}), 
which are generated using the default settings for the DES,
we can characterize this estimator for each photometric band because
the true time zero-point, $t_o$, of the simulations.  Note that for the SNANA
simulations, the true time of peak flux is computed in each band separately and 
then the results are combined using a filter-weighted average; thus, we are not
actually comparing our estimator, $\widehat{t}_o$, to the true time of {\it r}-band 
maximum, but rather to a slightly different, redshift-dependent quantity.}

\begin{figure}
\includegraphics[angle=-90,width=3.5in]{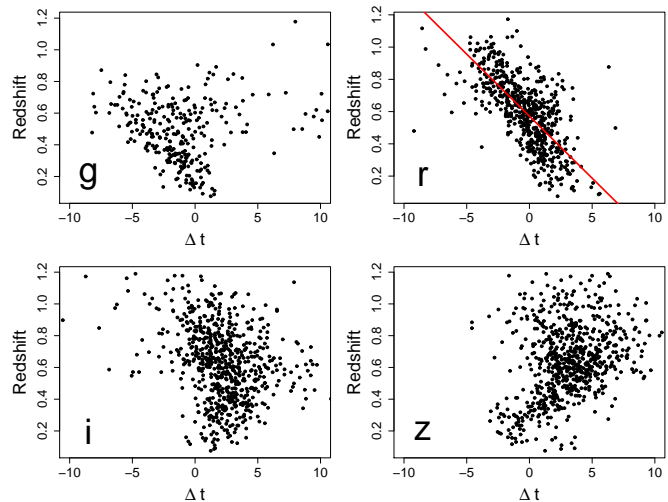}
\caption{
The difference $\Delta t = \hto - t_o$ (in days) versus redshift for
600 SNeIa simulated with {\tt SNANA} using its default settings for the
Dark Energy Survey.  The 600 SNe are the peaked SNe from a simulated sample of
1,000 SNe.
The top row shows $g$- and $r$-band data while
the bottom row shows $i$- and $z$-band data.
Overlaid on the $r$-band data is a linear regression fit to the data
in the range $-5 < \Delta t < 5$ that is meant to be purely illustrative.
}
\label{fig:dfilt}
\end{figure}

See Figure \ref{fig:dfilt} and Table \ref{tab:dfilt} for the results
for the 600 SNe in the sample with peaks (i.e., the regression spline
maximum is not at an end point).
The $x$-axis of Figure \ref{fig:dfilt} shows $\Delta t = \hto - t_o$,
while the second and third columns of Table \ref{tab:dfilt} show
the estimated mean and standard deviation for $\Delta t$.  (The fourth
column indicates the estimated correlation between $\Delta t$ and
redshift $z$.)
We find that using the time of the $r$-band peak flux as our
estimator $\hto$ is the proper choice: it is nearly
unbiased and it has the minimum variance.  

\begin{table}
\centering
\caption{Time Normalization Estimator: Different Filters. }
\begin{tabular}{@{}lccc@{}}
\hline
Filter & $\widehat{\mu}$ (days) & $\widehat{\sigma}$ (days) & $\widehat{\rho}$\\
\hline
g & 13.86 & 27.26 & 0.66 \\
r & -0.27 & 2.17 & -0.67 \\
i & 2.21 & 2.65 & -0.27 \\
z & 3.25 & 2.84 & 0.30\\
\hline
\label{tab:dfilt}
\end{tabular}
\end{table}

\begin{figure}
\includegraphics[angle=0,width=3.5in]{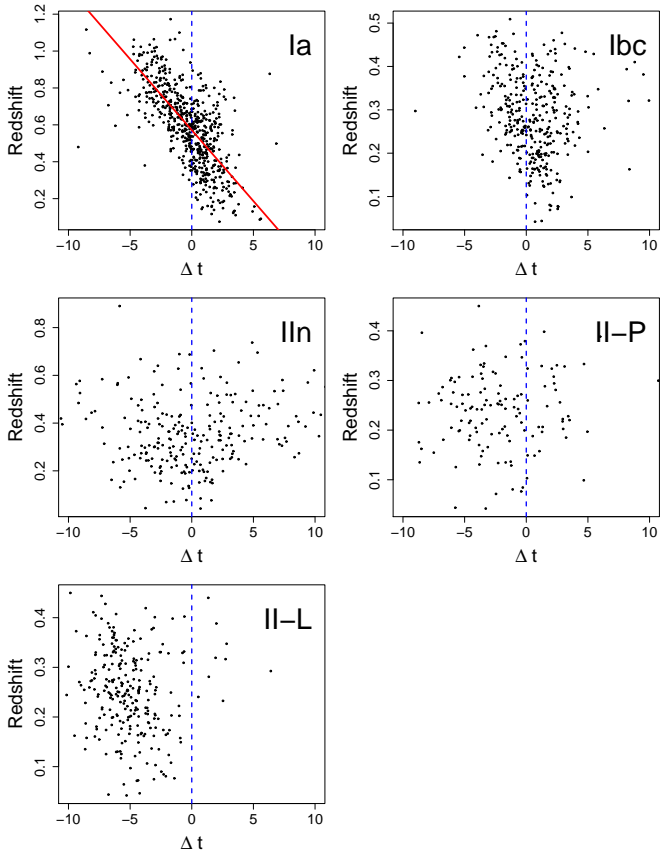}
\caption{
The difference $\Delta t = \hto - t_o$ (in days) versus redshift for
peaked SNeIa simulated with {\tt SNANA} using its default settings for the
Dark Energy Survey.
The top row shows $r$-band data for SN Types
Ia, Ibc, and IIn while the bottom row shows $r$-band data for SN Types
II-P and II-L.
Overlaid on the SNeIa results is a linear regression fit to the data
in the range $-5 < \Delta t < 5$ that is meant to be purely illustrative.
}
\label{fig:dsne}
\end{figure}

In Figure \ref{fig:dsne} and Table \ref{tab:dsne}, we characterize $\hto$,
the {\it r}-band max estimator of $t_0$,  given 1,000 examples each of different SN types,
each observed in the $r$ band.  The conclusion that we draw is that
if we corrected $\hto$ using host photo-$z$ estimates,
the effect on other SN types would be
minimal, judging by the standard deviations shown in Table \ref{tab:dsne}.

\begin{table}
\centering
\caption{Time Normalization Estimator: Different SN Types\label{tab:dsne}}
\begin{tabular}{@{}lccc@{}}
\hline
SN Type & $\widehat{\mu}$ (days) & $\widehat{\sigma}$ (days) & $\widehat{\rho}$\\
\hline
Ia & -0.27 & 2.17 & -0.67 \\
Ibc & 2.93 & 11.36 & -0.24 \\
IIn & 1.97 & 9.57 & -0.03 \\
II-P & 32.99 & 31.76 & -0.102 \\
II-L & -2.19 & 12.64 & -0.297\\
\hline
\end{tabular}
\end{table}

\begin{figure}
\includegraphics[angle=-90,width=3.5in]{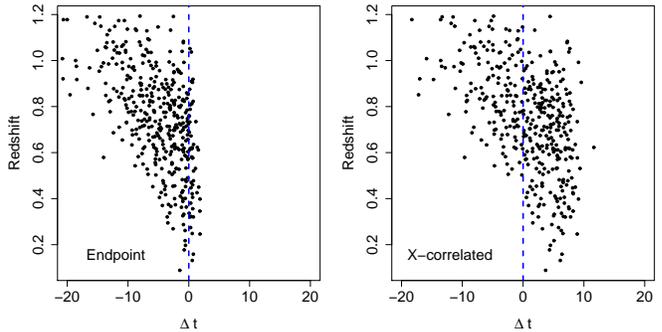}
\caption{
The difference $\Delta t = \hto - t_o$ (in days) versus redshift for
400 SNeIa simulated with {\tt SNANA} using its default settings for the
Dark Energy Survey.  The 400 SNe are the unpeaked SNe from a simulated
sample of 1,000 SNe.
The left panel shows the distribution of $\Delta t$ values assuming the
time of the first $r$-band datum for $\hto$, while the right panel shows
the same distribution except that $\hto$ is estimated by cross-correlating
the unpeaked light curve with all peaked light curves in the sample and
taking the mean (see eq.~\ref{eqn:nopeak}).
}
\label{fig:nopeak}
\end{figure}

Simulated SNe light curves without a peak in the {\it r}-band are treated differently from those with peaks.
To estimate $t_o$ for a given un-peaked light curve, we cross-correlate it with
with each SN that has an $r$-band peak.  This produces a sequence of 
estimates of $t_o$, $(\widehat{t_{o,j}})_{j=1}^M$, where $M$ is the number
of peaked SN.  Finally, we return 
\begin{equation}
\hto = \frac{1}{M} \sum_{j=1}^M \widehat{t}_{o,j} \,. \label{eqn:nopeak}
\end{equation}
as the estimate of $t_o$ for that unpeaked light curve.
To examine our use of cross-correlation, we return to the SNANA simulated 
SNe described above.    
In Figure \ref{fig:nopeak} we plot $\Delta t$ for using 
the endpoint and using cross-correlation for
each of the 400 unpeaked SN. 
Without cross-correlation, $(\widehat{\mu},\widehat{\sigma})$
= $(-5.02,4.76)$; with cross-correlation, the values are
$(0.87,5.63)$, i.e., a small increase in estimator standard deviation is offset
by a marked decrease in bias.  Thus we conclude that cross-correlation is an
effective approach.

\subsection{Light Curve Distance Metric}
\label{ss:lcdist}

Call the set of normalized light curves for supernova $i$ $\{\widetilde{t}_{ik}, \widetilde{F}_{ik}^b, \widetilde{\sigma}_{ik}^b\}$.  For each pair of supernovae, $\X_i$ and $\X_j$, we define the squared {\emph b-band} distance between them, where {\emph b} ranges over {\it griz}, as
\begin{equation}
\label{eqn:distb}
\label{eqn:distband}
s_b(\X_i,\X_j) = \frac{1}{t_u - t_l} \sqrt{\sum_{k: t_u \le k \le t_l} \frac{\left(\widetilde{F}_{ik}^b-\widetilde{F}_{jk}^b\right)^2}{(\widetilde{\sigma}^b_{ik})^2+(\widetilde{\sigma}^b_{jk})^2}}
\end{equation}
\new{where $k$ indexes the time grid of the smoothed supernova light curves; a time binning of 1 day is used.}
Hence, $s_b(\X_i,\X_j)$ is the weighted Euclidean distance between light curves, per overlapping time bin.  
The quantities $t_l=\max(\widetilde{t}_{i1},\widetilde{t}_{j1})$ and $t_u=\min(\widetilde{t}_{ip_i},\widetilde{t}_{jp_j})$ define the lower and upper time bounds, respectively, of the overlapping regions of the two light curves.  If two normalized light curves have no overlap, then their distance is set to a large reference value.  Finally, we define the \emph{total} distance between two light curves, as $s(\X_i,\X_j) = \sum_b s_b(\X_i,\X_j)$, the sum of the distances in eq.~(\ref{eqn:distb}), across bands.  This distance is used in eq.~(\ref{eqn:dmap1}) to build the weighted graph which we use to compute the diffusion map parametrization, $\{\mathbf{\Psi}_1,...,\mathbf{\Psi}_N\}$, of our data (equation \ref{eqn:dmap2}).
Each $\mathbf{\Psi}_i$ are the coordinates of the $i^{th}$ SN in the estimated diffusion space.

\section{Semi-Supervised Supernova Classification: The Supervised Part}
\label{sec:snapp1}

Once we have a parametrization, $\mathbf{\Psi}$, for each supernova, the next step
is to use a training set of SNe of known, spectroscopically confirmed type to learn a classification model to predict the type 
of each supernova from its representation $\mathbf{\Psi}$. This is the \emph{supervised} part
of our semi-supervised methodology.

\subsection{Constructing a Training Set}
\label{ss:trainset}

In the Supernova Challenge, the training set, $\mathcal{S}$, was generated assuming a
fixed amount of spectroscopic follow-up on each of a 4m and 8m-class telescope.
The magnitude limits were assumed to be 21.5 ($r$-band) for the 4m and 23.5 ($i$-band) for the 8m telescope (\citealt{kess2010a}).  

Using $\mathcal{S}$ as a training set is problematic for at least two reasons.  
First, $\mathcal{S}$ consists of SN that have much lower host-galaxy $z$ and higher observed brightness than those in $\mathcal{P}$ (see Fig. 2 in \citealt{kess2010a}).  Second, $\mathcal{S}$ oversamples the number of Type Ia SN relative to that in the entire data set, $\mathcal{D}$ (see Table \ref{tab:datasetsComposition}).  These distributional mismatches induce inadequate modeling in those parameter subspaces undersampled by $\mathcal{S}$ and can cause model selection procedures to choose poor models for classification in $\mathcal{P}$.  Both of these issues hinder attempts to generalize models fit on $\mathcal{S}$ to classify supernovae in $\mathcal{P}$. 

We study the dependence of the classification accuracy
on the particular training set employed (or more precisely, on the specific procedure used to perform spectroscopic follow-up).
In this section, we propose a variety of procedures to procure spectroscopically confirmed labels; in \S\ref{sec:results} we will analyse the sensitivity of our classification results to the training set used and
determine the optimal strategy for spectroscopic SN follow-up.

We phrase the problem in the following way: assuming that we have a fixed number of 
hours for spectroscopic follow-up, what is the optimal way to use that time?
To simplify our study, we assume that all follow-up is performed on an 8m telescope
with spectroscopic integration times given in Table \ref{tab:spectimes}, based on simplistic simulations under average conditions using the FOcal Reducer and low dispersion Spectrograph for the Very Large Telescope (VLT FORS) exposure
time calculator \footnote{\url{http://www.eso.org/observing/etc/bin/gen/form?INS.NAME=FORS2+INS.MODE=spectro}}.  The amount of
spectroscopic follow-up time necessary for each target is determined by the integration
time (Table \ref{tab:spectimes}) corresponding to its $r$-band maximum.\footnote{For non-integer magnitudes, the integration times are determined by a quadratic interpolation function.}  These integration times are meant to be approximate figures for means of constructing simulated spectroscopic training sets.

\begin{table}
\centering
\caption{Spectroscopic integration times assumed for follow-up observations of supernovae, based on average conditions at the VLT using the FORS instrument.}
\begin{tabular}{@{}cc@{}}
\hline
{$r$-band mag} & {integration time (minutes)}\\
\hline
   20 & 1\\
   21 & 2\\
   22 & 5\\
  23 & 20\\
  24 & 100\\
  25 & 600\\
 25.5 & 1500\\
\hline
\label{tab:spectimes}
\end{tabular}
\end{table}

Under this scenario, the spectroscopic data in $\mathcal{S}$ require a total of 24,000 minutes (400 hours) of integration time.
Assuming this amount of follow-up time\footnote{Note that we fix the total amount of \emph{integration} time needed, ignoring the time lost to overheads, weather, and other effects.}, we construct alternate training sets using each of the following procedures:
\begin{enumerate}
\item Observe SNe in order of decreasing brightness. This strategy observes only the brightest
objects, and allows us to obtain spectra for the maximal number of supernovae.  Call this $\mathcal{S}_B$.
\item Perform a ($r$-band) magnitude-limited survey, down to a prespecified magnitude cutoff.  Here, we
randomly sample objects brighter than the magnitude cut, until the observing time is filled.  We experiment 
with four different cutoffs: 23.5, 24, 24.5, and 25th magnitude.  Call these $\mathcal{S}_{m,23.5},\mathcal{S}_{m,24},\mathcal{S}_{m,24.5}$, and $\mathcal{S}_{m,25}$.
\item Perform a redshift-limited survey.  We try two different redshift cutoffs: $z$=0.4, 0.6.  Call these $\mathcal{S}_{z,0.4}$, and $\mathcal{S}_{z,0.6}$.
\end{enumerate}
The magnitude- and redshift-limited surveys both have a random component.  To quantify the effects that this randomness has on the samples and the ultimate supernova typing, we construct 15 data sets from each spectroscopic ``survey".  In Table \ref{tab:datasets}, we display the median number of SNe from each of $\mathcal{S}$ and $\mathcal{P}$ contained in each spectroscopic training set. Note that as the limits of the survey get fainter (and higher $z$), the ratio of elements from $\mathcal{P}$ to $\mathcal{S}$ increases, as the total number of SNe decreases.  Table \ref{tab:datasetsComposition} shows the median class composition of each training set.  Here, the deeper training sets more closely resemble the class distribution in $\mathcal{D}$. We will return to these training sets in \S\ref{sec:results}.

\begin{table*}
\centering
\caption{Composition of the training datasets, broken down by the SN Challenge spectroscopic/photometric designation.  In the first two rows, each cell's entry shows the number of elements in the set that is the intersection of  $\mathcal{S}$ or $\mathcal{P}$ with the respective training set.  In the third row, the total number of objects in each training set is given.}
\begin{tabular}{@{}c|rrrrrrrrr@{}}
\hline
{Set} & {$\mathcal{S}$} & {$\mathcal{S}_{B}$}   &  {$\mathcal{S}_{m,23.5}$}&  {$\mathcal{S}_{m,24}$} & {$\mathcal{S}_{m,24.5}$}   & {$\mathcal{S}_{m,25}$} &  {$\mathcal{S}_{z,0.4}$} &  {$\mathcal{S}_{z,0.6}$} & {$\mathcal{D}$}  \\
\hline 
$\mathcal{S}$ & 1103 &      686 &  294 &  73 &  26 &  11 &  44 &  15 &  1103\\
$\mathcal{P}$ &     0 &     1765 &  979 & 508 & 272 & 155 & 240 & 117 & 19792\\
Total &    1103 &     2451 & 1273 & 587 & 302 & 165 & 284 & 135 & 20895\\
\hline
\label{tab:datasets}
\end{tabular}

Median values are displayed over 15 training sets.  Only for $\mathcal{S}$ and $\mathcal{S}_B$ are the training set identical on each iteration.
\end{table*}

\begin{table*}
\centering
\caption{Composition of the training datasets, broken down by the SN Challenge spectroscopic/photometric designation.}
\begin{tabular}{@{}c|rrrrrrrrr@{}}
\hline
{SN Type} & {$\mathcal{S}$} & {$\mathcal{S}_{B}$}   &  {$\mathcal{S}_{m,23.5}$}&  {$\mathcal{S}_{m,24}$} &  {$\mathcal{S}_{m,24.5}$} &  {$\mathcal{S}_{m,25}$} &  {$\mathcal{S}_{z,0.4}$} &  {$\mathcal{S}_{z,0.6}$} & {$\mathcal{D}$}  \\
\hline
Ia  &  559 & 1313 &  557 &  178 &   75 &   39 &   45 &   22 &  5088\\
Ib/c &    15 &   18 &   13 &    5 &    4 &    2 &    7 &    1 &   259\\
Ib &   71  & 168 &   79 &   32 &   15 &    8 &   30 &   10 &  1438\\
Ic &   58 &   88 &   50 &   26 &   13 &    8 &   36 &   14 &  1104\\
IIn &   63 &  131 &   84 &   42 &   24 &   14 &   10 &    5 &  1939\\
II-P &  326 &  707 &  480 &  295 &  161 &   90 &  159 &   74 & 10642\\
II-L &   11 &   26 &   12 &    5 &    3 &    2 &   17 &    6 &   425\\
\hline
\label{tab:datasetsComposition}
\end{tabular}

Median values are displayed over 15 training sets.  Only for $\mathcal{S}$ and $\mathcal{S}_B$ are the training set identical on each iteration.
\end{table*}

\subsection{Tuning the Classifier}
\label{ss:classtune}

We build a random forest classifier, $\widehat{h}$, by training on the diffusion map representation and known type
of a set of spectroscopically confirmed SNe.  This classifier then allows us to predict the class of each newly observed SN
light curve. In constructing such a classifier, there
are three tuning parameters which must be chosen:
\begin{enumerate}
\item $\epsilon$, the diffusion map bandwidth in eq.~(\ref{eqn:dmap1}),
\item $m$, the number of diffusion map coordinates used in the classifier, and
\item $\tclass$, the minimum proportion of random forest trees predicting that a SN is of Type Ia necessary for us to decide that the SN is Type Ia.
\end{enumerate}
In this section we describe how to choose these parameters in a statistically rigorous way.

 The SN Classification Challenge was based on correctly predicting the Type Ia supernovae.  The Challenge used the Type Ia Figure of Merit (FoM)
\begin{equation}
\label{eqn:fom}
\hf = \frac{1}{N_{\rm Ia}^{\rm Total}} \frac{(N_{\rm Ia}^{\rm true})^2}{N_{\rm Ia}^{\rm true} + WN_{\rm Ia}^{\rm false}}
\end{equation}
where $N_{\rm Ia}^{\rm Total}$ is the total number of Type Ia 
SNe in the sample, $N_{\rm Ia}^{\rm true}$ is the number of Type Ia SNe 
correctly predicted, and $N_{\rm Ia}^{\rm false}$ is the number of non-Ia SNe 
incorrectly predicted to be Type Ia.  The factor $W$ controls the relative 
penalty on false positives over false negatives.  For the SN Photometric 
Classification Challenge, $W\equiv 3$.  This penalty on Type Ia purity means 
that we need to be conservative in calling a SN a Type Ia: we are 
penalized three times more by calling a non-Ia a Type Ia than in calling a 
Type Ia a non-Ia.  

Since the Challenge gives an explicit criterion (equation \ref{eqn:fom}), we choose the tuning parameters that 
maximize $\hf$.  To avoid overfitting to the training set,
we maximize a 10-fold cross-validation estimate of $\hf$ and call this maximum value $f^*_{\rm Ia}$.  We find that $f^*_{\rm Ia}$
is insensitive to the value of $m$ for a large enough $m$, as the random forest largely ignores
irrelevant components, so for the rest of the analysis we fix $m = 120$. 
To find the optimal model, $(\estar,\tstar)$, we perform a two dimensional grid search over $(\epsilon,\tclass)$.
Once this optimal model is discovered by minimizing the cross-validated
$\hf$, it is applied to the photometric sample to predict the class of each supernova.

\subsection{Incorporating Redshift Information}
\label{ss:redshiftDescription}


%
%
%
In addition to observed light curves, host-galaxy (photometric) redshift estimates, 
$z$, are often available.  If this information is known, it should be included in
the supernova analysis.
We may incorporate redshift information in one of two ways:
\begin{itemize}
\item directly in the calculation of the pairwise distance metric, $s$; or
\item as an additional covariate for the classification model, $h$.
\end{itemize}
In the former approach, we artificially inflate the distance measure
between two SNe, $i$ and $j$, if
\begin{equation}
\label{eqn:hostz}
\frac{\vert z_i - z_j \vert}{\sqrt{u_i^2 + u_j^2}} > n_s \,,
\end{equation}
where $u$ denotes the estimated redshift uncertainty, and $n_s$ is
a set constant (e.g., 3).  Using eq. (\ref{eqn:hostz}), we effectively
deem two supernovae `dissimilar' if their redshifts are greater than $n_s$
standard deviations apart, even if their light curve distance, $s$, is small.
This approach can aid to break redshift degeneracies in light curve shapes
 and colours.

In the latter approach, we simply use $z$ as a covariate in the classification model
in addition to the diffusion map coordinates.  This treatment of redshift allows
the classification model to directly learn the dependencies of supernova type on redshift.
However, this method relies on the assumption that the training SN distribution
resembles that of the set to be classified (and, crucially, that the entire range of
redshifts is captured by the training data).  The first approach does not rely on
such assumptions because the redshifts are employed, in an unsupervised manner,
to alter the SN representation prior to learning a classification model.

In Tables \ref{tab:zclassIa} and \ref{tab:zclassIIp}, we show the classification
results, on the SN Challenge data, of incorporating redshift into our analysis using
each of the two methods.  The strategy of using redshifts to directly influence the diffusion map
coordinates performs better than including the redshifts as a covariate in the 
classification model, and both methods of incorporating redshift information provide improved 
classifications (see \S\ref{ss:zres}).

%
%
\subsection{Computational Resources}
Though fitting the above described method can take substantial computational resources, the splines,
spectral decomposition, and random forest have very stable and efficient implementations that 
are readily used and highly optimized.  The dominating computation remains computing the distance matrix,
with the bottleneck being the cross-correlation approach to estimating the time of maximum flux of a non-peaked
SN light curve.

\section{Results}
\label{sec:results}

In this section, we analyse 20,895 supernovae provided in the Supernova
Photometric Classification Challenge using the methods introduced in \S\S \ref{sec:meth}-\ref{sec:snapp1}.

\subsection{Diffusion Map Representation of SNe}

\begin{figure}
\includegraphics[angle=-90,width=3.5in]{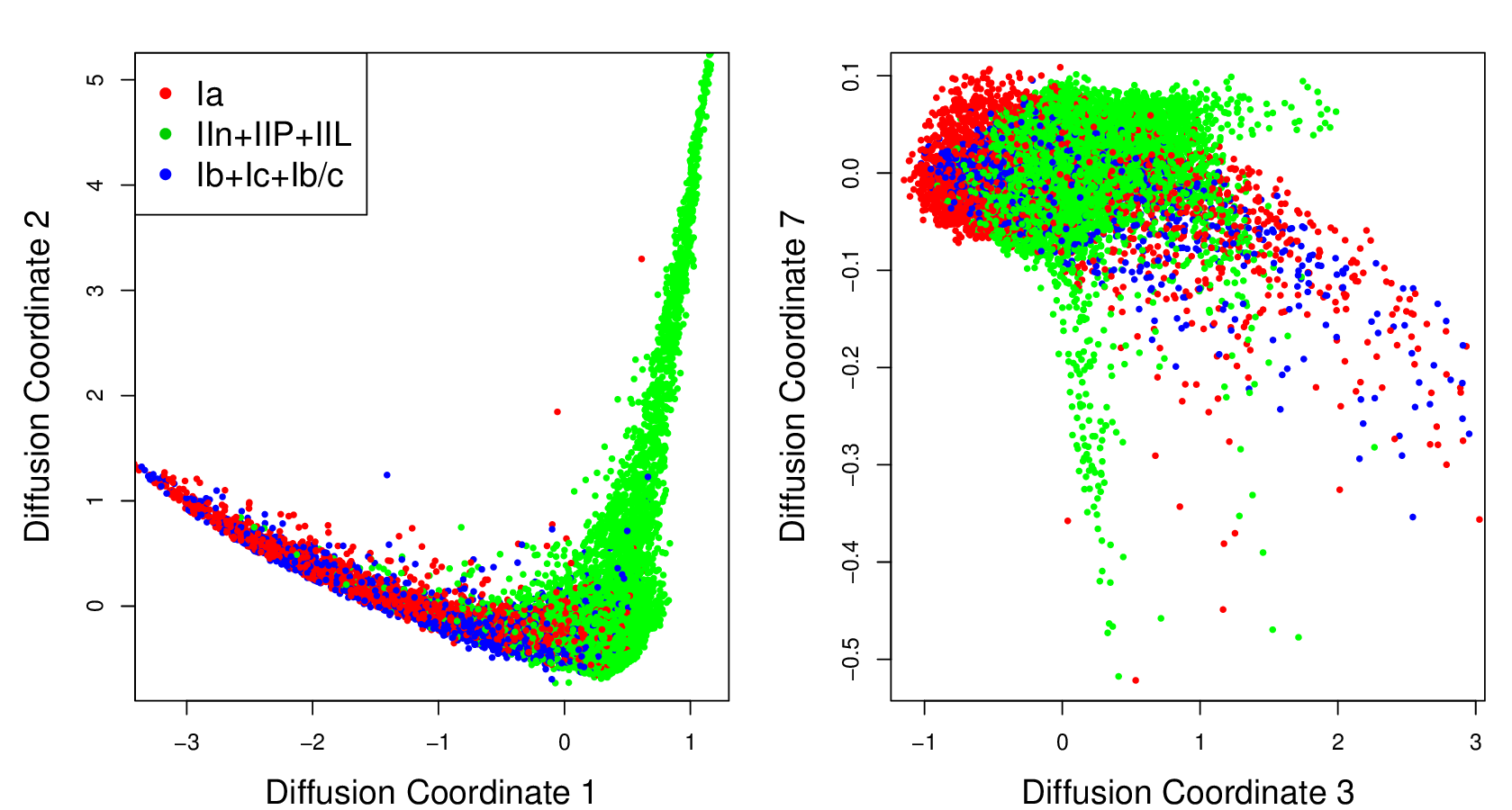}\\ 
\includegraphics[angle=-90,width=3.5in]{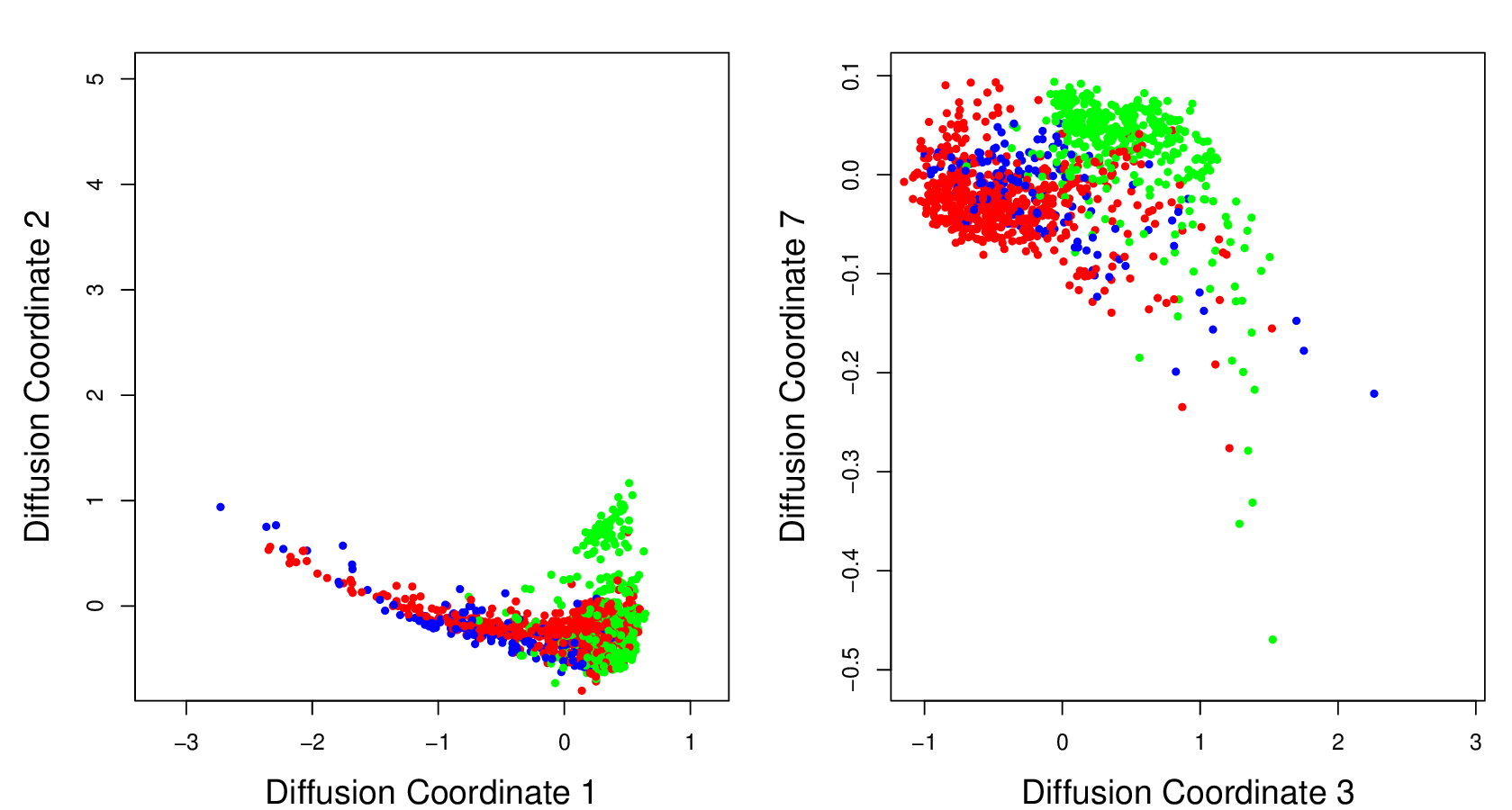}
\caption{
Top: Diffusion coordinates for all (spectroscopic+photometric) SNe, of Type Ia (red), IIn+II-P+II-L (green),
and Ibc+Ib+Ic (blue).  Bottom: Diffusion coordinates for only the spectroscopic SN sample.  Left: In the first two
coordinates, there is clear separation between the Type II and Type I SNe.  Right: Plotted are the two most important
diffusion coordinates for SN classification, as determined by a random forest classifier built on the training set.  Clear
separations between SN types in the spectroscopic set are not present in the combined set.}
\label{fig:impnoz}
\end{figure}

The first step in this analysis is to compute the diffusion map representation
of each SN.  In Figure \ref{fig:impnoz} we plot diffusion map coordinates,
coloured by true class, for the entire SN set, $\mathcal{D}$ (top) and spectroscopic set, $\mathcal{S}$ (bottom).
These coordinates were computed without using host photometric redshifts.
Large discrepancies between the diffusion map distributions of these two sets
are obvious.  The left panels of Figure \ref{fig:impnoz} show the first two
diffusion coordinates, where clear separation between the Type I and Type II SNe
is obvious.  The right panels plot the two most important coordinates (3 \& 7), as
estimated by a random forest classifier trained only on the spectroscopic data.
With respect to these two coordinates, there is 
separation between the Type Ia and Ibc supernovae in the spectroscopic set. 

\begin{figure}
\includegraphics[angle=0,width=3.5in]{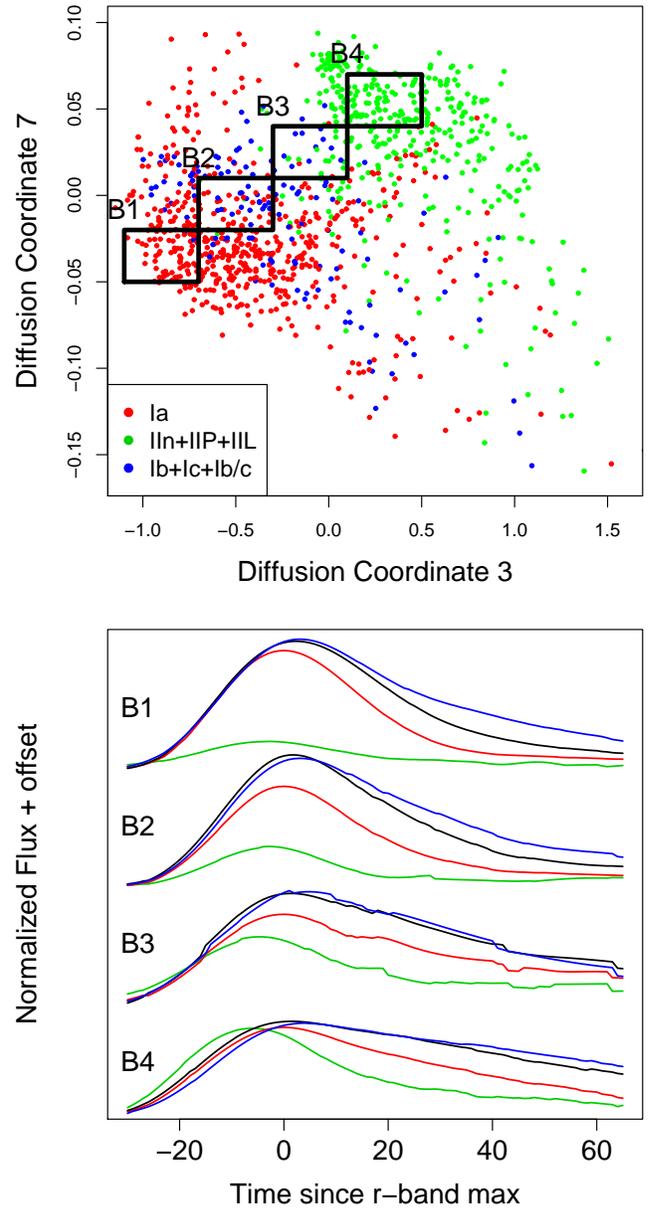}\\ 
\caption{
Top: Supernovae in the spectroscopic sample $\mathcal{S}$ show a separation between the various supernova types in diffusion coordinates 3 and 7.  Bottom:  The average $griz$ light curves in each of the four boxes B1-B4 are plotted, revealing a gradual flattening of the light curves and an incremental increase in the relative strength of the $g$-band emission. }
\label{fig:lcdmap}
\end{figure}

In Fig. \ref{fig:lcdmap}, we plot the average 4-band light curves of the SNe within each of 4 regions in diffusion space.
Stepping across this coordinate system (with respect to coordinates 3 and 7), we see that the relative strength of the 
$g$-band emission increases and the light curves become more plateaued.  This corresponds to a gradual transition from 
Type Ia to Type II SNe.

\begin{figure}
\includegraphics[angle=-90,width=3.5in]{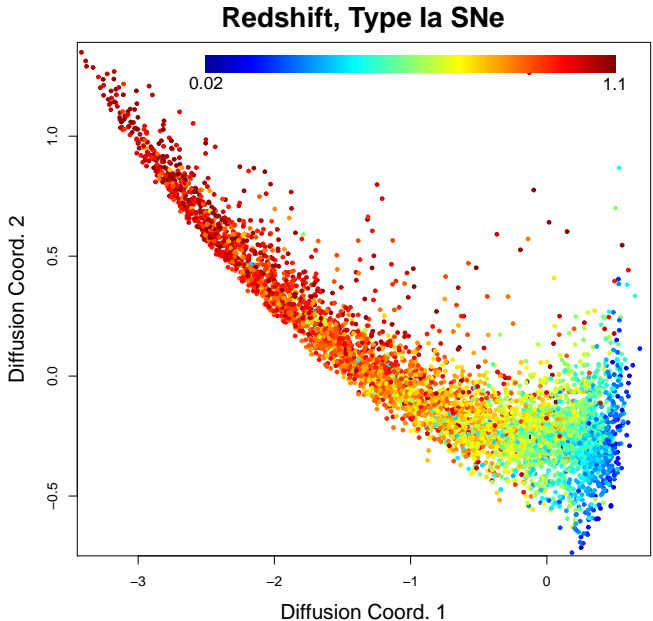}
\caption{
Redshift of all 5088 Type Ia supernovae, plotted in the first two diffusion map coordinates.  The true redshift varies gradually across
diffusion space, even though this information was not used to construct the diffusion map.}
\label{fig:zdmap}
\end{figure}

We also explore the behavior, in diffusion space, of SN redshift.  
Fig. \ref{fig:zdmap} plots the first two diffusion coordinates of the 5088 Ia SNe,
coloured by their true redshift.  Even though we did not use any redshift information
in the computation of these coordinates, we see a gradual trend in redshift 
with respect to this coordinate system, with an increase in 
$z$ as one moves diagonally from bottom right to top left.  
This means that our construction of distance measure for the diffusion 
map captures the incremental changes with respect to redshift that occur 
in the light curves.   Note that using the entire set of data to estimate the diffusion 
map coordinates has allowed us a fine enough sampling with respect to redshift;
this would generally not be the case if we were to employ only the spectroscopic SNe
to build the diffusion map.  This is a critical result because it shows that the 
main source of variability in diffusion space is due to SN type, and not
redshift.  
Hence, we can use the diffusion map representation of the SNe directly,
without having to estimate the redshift of each SN and correct for its effects
on the light curves.  In \S\ref{ss:zres} we will compare this result with using
 host-galaxy photo-z estimates to directly alter the diffusion map coordinates.

\subsection{Classification of Type Ia SNe}

Given the diffusion map representation of the entire set of SNe, $\mathcal{D}$ (Fig. \ref{fig:impnoz}, top)
and a training set (see \S\ref{ss:trainset}), we train a classification model
to predict the types of the remaining supernovae.  To achieve this,
we tune the classifier on our training set, as in \S\ref{ss:classtune}, and then
apply this classifier to $\mathcal{P}$.

In Table \ref{tab:classIa} we display the results of Type Ia
SN classification with the eight training datasets described in \S\ref{ss:trainset}.
For each training set, we report the optimal training set Type-Ia FoM, $f^*_{\rm Ia}$, and tuning parameters, 
$(\estar,\tstar)$, and the Type-Ia FoM, $\widehat{f}_{\rm Ia, pred}$, of the predictions for $\mathcal{P}$.\footnote{For each training set, we evaluate $\widehat{f}_{\rm Ia, pred}$ for only those points in $\mathcal{P}$ not included in the training set.}  Additionally, we show the purity and
efficiency of Type-Ia classification, defined as
\begin{eqnarray}
\hp = \frac{N_{\rm Ia}^{\rm true}}{N_{\rm Ia}^{\rm true} + N_{\rm Ia}^{\rm false}}
\end{eqnarray}
and
\begin{eqnarray}
\he = \frac{N_{\rm Ia}^{\rm true}}{N_{\rm Ia}^{\rm Total}}.
\end{eqnarray}
Note that all entries in Table \ref{tab:classIa} are median values over 15 repetitions of
each training set.\footnote{Training sets $\mathcal{S}$ and $\mathcal{S}_B$ are the same
for each iteration, but results differ slightly on each iteration due to randomness in the
random forest classifier.}

Results of this experiment show that the deeper magnitude-limited follow-up strategies perform
the best, achieving a $\widehat{f}_{\rm Ia, pred}$ of 0.305, a value 2.4 times the FoM achieved by the classifier trained
on $\mathcal{S}$.  For the $\mathcal{S}_{m,25}$ training procedure,
$(\widehat{p}_{\rm Ia, pred},\widehat{e}_{\rm Ia, pred})=(0.72,0.65)$.  For each training set, the cross-validated
Type Ia figure of merit estimated on the training data is significantly larger than the figure of merit achieved
on the photometric data, showing that none of the training sets completely resembles $\mathcal{P}$.  Notably, on the Challenge
training set, $\mathcal{S}$, our method achieves a cross-validated Type Ia purity/efficiency
of 95\%/87\%, but this transfers to a purity/efficiency of 50\%/50\% on the data in $\mathcal{P}$. 

\begin{figure}
\includegraphics[angle=0,width=3.25in]{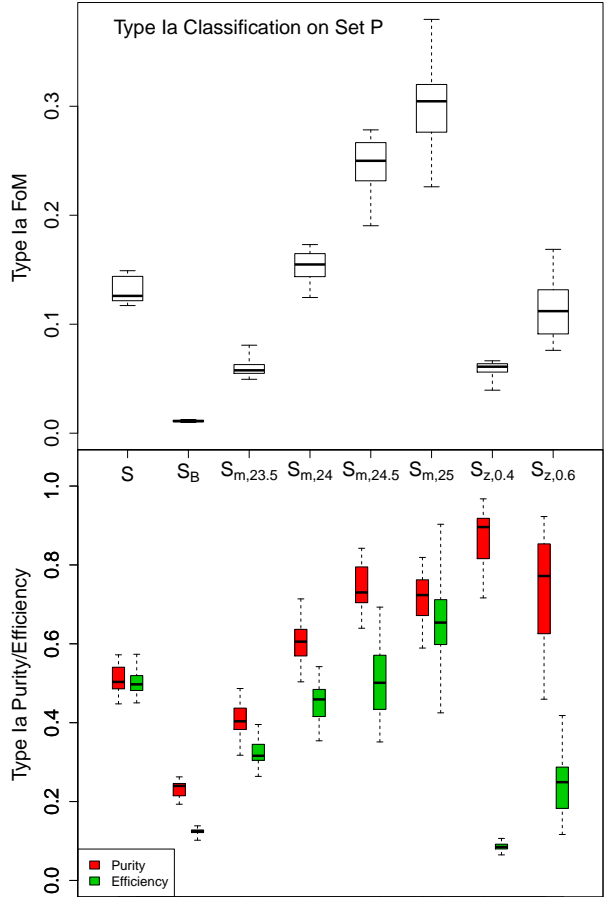}
\caption{ Performance on Type Ia SNe in $\mathcal{P}$ of classifiers trained on spectroscopically confirmed supernovae
from 8 different follow-up procedures.  Top: Type Ia Figure of Merit, $\widehat{f}_{\rm Ia, pred}$.  Bottom: Type Ia
purity and efficiency.  Boxplots show the distribution of each metric over 15 training sets.  The obvious winner
is the deeper magnitude-limited survey, which achieves median purity and efficiency between 65-75\%.}
\label{fig:Iaphot}
\end{figure}

Figure \ref{fig:Iaphot} displays the distribution of Type Ia FoM, purity, and efficiency for each of the spectroscopic follow-up
procedures.  A few observations:
\begin{itemize}
\item The deeper magnitude-limited surveys perform the best in terms of FoM.  The obvious trend is that deeper surveys perform better, even though their training sets are much smaller.  For instance, the 23.5 mag. limited
survey contains, on average, 7.7 times the number of training SNe as the 25 magnitude limited survey, but attains a prediction FoM a mere 19\% as large.
\item Compared to the Challenge training set, $\mathcal{S}$, the 25th magnitude-limited survey attains a 44\% increase in purity and 30\% increase in 
efficiency of Type Ia SNe.
\item The worst strategy is to follow-up on only the brightest SNe.  Though this maximizes the number of labelled supernovae, it produces
the smallest figure of merit.
\item Redshift limited surveys are suboptimal to magnitude-limited surveys.  Though this strategy results in Type Ia samples with high purity, the efficiency of these samples  is very low.  A redshift-limited survey is not ideal for our approach because we have not directly modeled the redshift dependence of SN light curves. Without the ability to capture high $z$ SNe, our model becomes overly conservative, resulting in low efficiency.
\item Though there can be large variability in the actual samples produced by the magnitude-limited surveys, the Type Ia FoM is very stable, with FoM interquartile range typically smaller than 0.05.
\end{itemize}
Based on this study, we recommend that deeper magnitude-limited follow-up strategies be used to attain SN training samples.  Using a 25th magnitude-limited follow-up procedure yields a Ia FoM of 242\% that of the shallower procedure used in the SN Challenge.

\begin{table}
\centering
\caption{Results of Classifying Type Ia Supernovae using training sets from \S\ref{ss:trainset}.\label{tab:classIa}}
\begin{tabular}{@{}l|cccccc@{}}
\hline
Training Set & $\estar$ & $\tstar$ & $f^*_{\rm Ia}$ & $\widehat{f}_{\rm Ia, pred}$ & $\widehat{p}_{\rm Ia, pred}$ & $\widehat{e}_{\rm Ia, pred}$ \\
\hline
$\mathcal{S}$ & 1.4 & 0.58 & 0.757 & 0.126 & 0.503 & 0.497\\
$\mathcal{S}_B$ & 1.4 & 0.65 & 0.840 & 0.011 & 0.240 & 0.125\\
$\mathcal{S}_{m,23.5}$ & 1.4 & 0.54 & 0.796 & 0.058 & 0.404 & 0.316\\
$\mathcal{S}_{m,24}$ & 1.0 & 0.46 & 0.728 & 0.155 & 0.605 & 0.459\\
$\mathcal{S}_{m,24.5}$ & 1.0 & 0.45 & 0.610 & 0.250 & 0.730 & 0.501\\
$\mathcal{S}_{m,25}$ & 1.0 & 0.37 & 0.494 & {\bf 0.305} & 0.724 & 0.654\\
$\mathcal{S}_{z,0.4}$ & 1.4 & 0.41 & 0.664 & 0.061 & 0.896 & 0.085\\
$\mathcal{S}_{z,0.6}$ & 1.2 & 0.29 & 0.600 & 0.112 & 0.772 & 0.249\\
\hline
\end{tabular}
$f^*_{\rm Ia}$ is computed on training set via 10-fold cross-validation.\\
$\widehat{f}_{\rm Ia, pred}$, $\widehat{p}_{\rm Ia, pred}$, and $\widehat{e}_{\rm Ia, pred}$ are evaluated on all data in the photometric set $\mathcal{P}$ not in the training set.
\end{table}

\subsection{Type II-P Classification}

Type II-P supernovae are also useful for cosmological studies because they can be used as standard candles (\citealt{pozn2010}).
Here, we classify Type II-P supernovae in the SN Challenge data set using the above methods and
analogous Type II-P figure of merit.

\begin{figure}
\includegraphics[angle=0,width=3.25in]{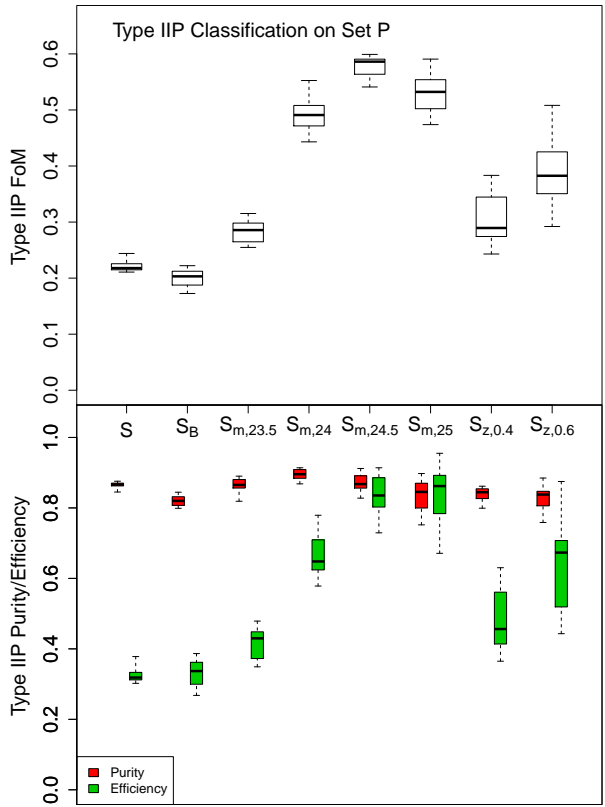}
\caption{ Same as Figure \ref{fig:Iaphot} for Type II-P classification.  Here, a 24.5th magnitude-limited survey attains
the maximal II-P figure of merit. Each spectroscopic training strategy results in a high II-P purity, but a much different classification efficiency.
}
\label{fig:IIpphot}
\end{figure}

In Table \ref{tab:classIIp} we display the results of Type II-P supernova classification with each of the 8 training sets, and in
 Fig. \ref{fig:IIpphot} we plot the distribution of the Type II-P FoM, purity and efficiency with respect to each spectroscopic follow-up strategy.
 Much like the Ia study, we find that the deeper magnitude-limited surveys perform the best.  We find that in terms of Type II-P
 figure of merit, a 24.5th magnitude limited survey performs the best, achieving $\widehat{f}_{\rm IIP, pred}=0.586$, with purity and efficiency
 at 87\% and 84\%, respectively.  
 
 Qualitatively, the Type II-P figures of merit resemble those of the type Ia study.  For each training set, the purity of the classifications is quite high,
  above 80\%, while the efficiency differs greatly, from a minimum of 32\% for $\mathcal{S}$ to a maximum of 86\% for $\mathcal{S}_{m,25}$.  Compared
  to the training set, $\mathcal{S}$, used in the SN Challenge, a 24.5th magnitude-limited survey achieves only a slightly better II-P purity, but a 1.6 times increase
  in II-P efficiency.

\begin{table}
\centering
\caption{Results of Classifying Type II-P Supernovae using training sets from \S\ref{ss:trainset}.\label{tab:classIIp}}
\begin{tabular}{@{}l|cccccc@{}}
\hline
Training Set & $\estar$ & $t^*_{\rm IIP}$ & $f^*_{\rm IIP}$ & $\widehat{f}_{\rm IIP, pred}$ & $\widehat{p}_{\rm IIP, pred}$ & $\widehat{e}_{\rm IIP, pred}$ \\
\hline
$\mathcal{S}$ & 1.6 &0.55 &0.834 &0.218 &0.866 &0.319\\
$\mathcal{S}_B$ & 1.6 &0.49 &0.835 &0.203 &0.820 &0.337\\
$\mathcal{S}_{m,23.5}$ & 1.6 &0.54 &0.826 &0.286 &0.865 &0.430\\
$\mathcal{S}_{m,24}$  &1.2 &0.59 &0.791 &0.491 &0.896 &0.648\\
$\mathcal{S}_{m,24.5}$  &1.0 &0.52 &0.747 &{\bf 0.586} &0.868 &0.835\\
$\mathcal{S}_{m,25}$  &1.0 &0.48 &0.593 &0.532 &0.845 &0.862\\
$\mathcal{S}_{z,0.4}$  &1.4 &0.57 &0.745 &0.289 &0.844 &0.456\\
$\mathcal{S}_{z,0.6}$  &1.2 &0.55 &0.660 &0.383 &0.838 &0.673\\
\hline
\end{tabular}
$f^*_{\rm IIP}$ is computed on training set via 10-fold cross-validation.\\
$\widehat{f}_{\rm IIP, pred}$, $\widehat{p}_{\rm IIP, pred}$, and $\widehat{e}_{\rm IIP, pred}$ are evaluated on all data in the photometric set $\mathcal{P}$ not in the training set.
\end{table}

\subsection{Incorporating Host Redshift}
\label{ss:zres}

Finally, we study the performance of the two methods of incorporating host-galaxy redshifts (\S\ref{ss:redshiftDescription}).  In Tables \ref{tab:zclassIa} and
\ref{tab:zclassIIp} we show the results of classifying Type Ia and II-P SNe, respectively, using each of the two strategies for incorporating redshifts.  Results are shown for each of the 8 training sets, where the optimal redshift cutoff, $n_s$ in eq. (\ref{eqn:hostz}), was chosen by maximizing the cross-validated training set FoM over a grid of integer values from 2 to 6.

There is a clear improvement to the FoMs by including host-galaxy redshifts.  Compared to the non-redshift results in Tables \ref{tab:classIa}-\ref{tab:classIIp}, the FoM values  increase for every training set by the use of redshift to alter the diffusion map coordinates.  Using redshifts to alter the diffusion map coordinates consistently performs better than using redshift as a covariate.  Overall, the best strategy for including host-galaxy redshifts for Type Ia classification is to use eq. (\ref{eqn:hostz}) with $n_s=2$ on the training set $\mathcal{S}_{m,25}$.  Using this prescription yields a Type Ia FoM of 0.355, an improvement of 16\% over the best FoM without redshift.  For Type II-P classification, the best strategy is to use $n_s=4$, resulting in a FoM of 0.612, which represents an improvement of 4.4\%.

\begin{figure}
\includegraphics[angle=0,width=3.25in]{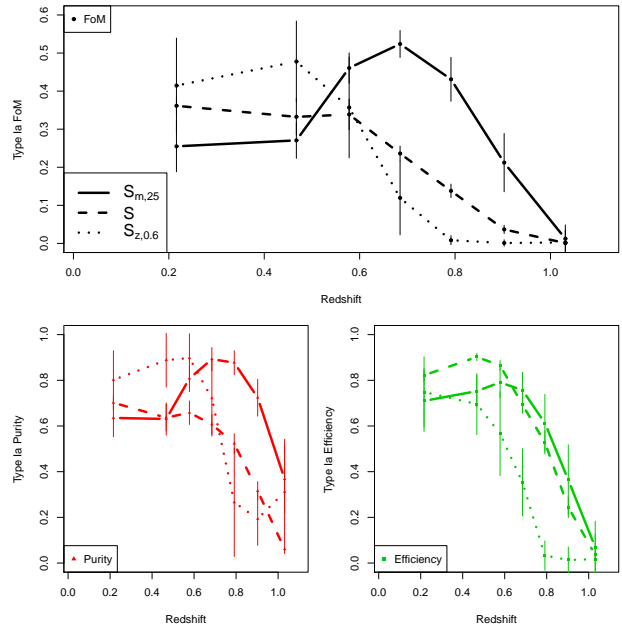}
\caption{ Performance of Type Ia classification as a function of redshift, for classification without using host-galaxy photometric redshifts.  In 7 equal sized redshift bins, the median FoM (top), purity (bottom left), and efficiency (bottom right) are plotted for three different training sets.  All three training sets result in poor estimates for high redshift SNe.  The magnitude-limited training set $\mathcal{S}_{m,25}$ (solid line) performs poorly for low redshifts, but the best for redshifts $> \sim 0.5$. The Challenge training set ($\mathcal{S}$, dashed line) performs poorly for all redshifts, while the redshift-limited training set $\mathcal{S}_{z,0.6}$ (dotted line)  performs the best for low redshifts but declines sharply to zero after reaching its redshift limit.}
\label{fig:fomz}
\end{figure}

\begin{figure}
\includegraphics[angle=0,width=3.25in]{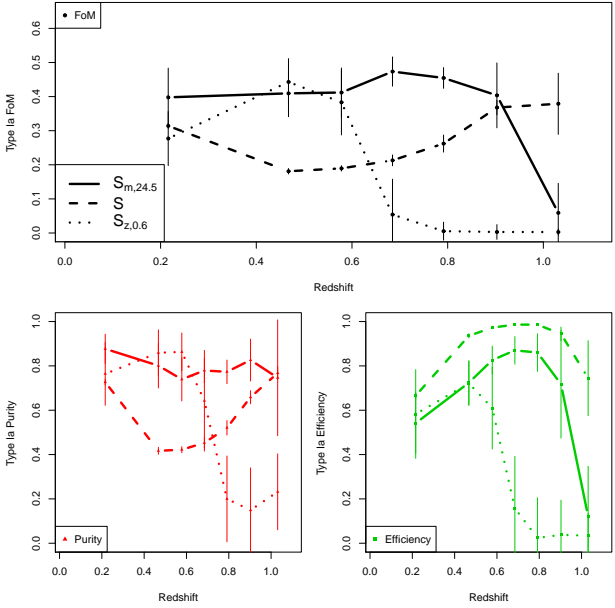}
\caption{ Same as Figure \ref{fig:fomz}, using photometric redshifts to alter the diffusion map coordinates, with $n_s=2$.   The magnitude-limited training set $\mathcal{S}_{m,25}$ (solid line) performs similarly across all redshifts except the highest redshift bin, besting the figure of merit for the Challenge training set ($\mathcal{S}$, dashed line) for six of the seven redshift bins.  Performance for the redshift-limited training set $\mathcal{S}_{z,0.6}$ (dotted line) declines sharply after reaching the redshift limit.}
\label{fig:fomz4}
\end{figure}

We plot the Type Ia FoM, purity, and efficiency as a function of redshift in Figure \ref{fig:fomz} for the analysis without using photometric redshifts, and in Figure \ref{fig:fomz4} for the analysis incorporating photo-z's with $n_s=2$.  Within each of 7 equally sized redshift bins, the median performance measures are plotted.  The magnitude-limited  training set experiences improvements in performance for both the lowest and highest redshifts.  The purity of the Type Ia sample found using the magnitude-limited training sample improves significantly at the lowest and highest redshifts by incorporating redshifts.  This highlights the tremendous value that host redshifts have in Type Ia classification, especially at breaking degeneracies between supernova type and redshift.

\begin{table}
\centering
\caption{Results of Classifying Type Ia Supernovae incorporating host redshifts. \label{tab:zclassIa}}
\begin{tabular}{@{}l|ccccccc@{}}
\hline
Tr. Set & $n^*_s$ & $\estar$ & $\tstar$ & $f^*_{\rm Ia}$ & $\widehat{f}_{\rm Ia, pred}$ & $\widehat{p}_{\rm Ia, pred}$ & $\widehat{e}_{\rm Ia, pred}$ \\
\hline
$\mathcal{S}$ &2 & 1.1 & 0.53 & 0.871 & 0.249 & 0.539 & 0.9\\
&-- & 1.2 & 0.59 & 0.84 & 0.131 & 0.446 & 0.584\\
$\mathcal{S}_B$ &6 & 1.2 & 0.55 & 0.903 & 0.029 & 0.24 & 0.308\\
&-- & 1.6 & 0.58 & 0.899 & 0.014 & 0.18 & 0.213\\
$\mathcal{S}_{m,23.5}$ &6 & 1.4 & 0.53 & 0.873 & 0.108 & 0.463 & 0.482\\
&-- & 1.4 & 0.57 & 0.848 & 0.06 & 0.369 & 0.37\\
$\mathcal{S}_{m,24}$ &4 & 1 & 0.47 & 0.799 & 0.252 & 0.726 & 0.564\\
&-- & 1.4 & 0.51 & 0.733 & 0.153 & 0.633 & 0.463\\
$\mathcal{S}_{m,24.5}$ &5 & 1.2 & 0.44 & 0.689 & 0.315 & 0.769 & 0.594\\
&-- & 1 & 0.43 & 0.649 & 0.232 & 0.716 & 0.503\\
$\mathcal{S}_{m,25}$ &2 & 1.2 & 0.39 & 0.615 & {\bf 0.355} & 0.758 & 0.741\\
&-- & 1 & 0.39 & 0.54 & 0.308 & 0.732 & 0.688\\
$\mathcal{S}_{z,0.4}$ &4 & 1.1 & 0.45 & 0.649 & 0.065 & 0.923 & 0.078\\
&-- & 1.6 & 0.41 & 0.671 & 0.058 & 0.87 & 0.084\\
$\mathcal{S}_{z,0.6}$ &4 & 1.4 & 0.39 & 0.562 & 0.104 & 0.831 & 0.206\\
&-- & 1.6 & 0.32 & 0.591 & 0.116 & 0.761 & 0.257\\
\hline
\end{tabular}

$n_s=$ -- indicates that host-galaxy redshift is used as a covariate in the Random Forest classifier, and not to construct diffusion map.\\
$f^*_{\rm Ia}$ is computed on training set via 10-fold cross-validation.\\
$\widehat{f}_{\rm Ia, pred}$, $\widehat{p}_{\rm Ia, pred}$, and $\widehat{e}_{\rm Ia, pred}$ are evaluated on all data in the photometric set $\mathcal{P}$ not in the training set.
\end{table}

\begin{table}
\centering
\caption{Results of Classifying Type II-P Supernovae incorporating host redshifts. \label{tab:zclassIIp}}
\begin{tabular}{@{}l|ccccccc@{}}
\hline
Tr. Set & $n^*_s$ & $\estar$ & $\tstar$ & $f^*_{\rm IIP}$ & $\widehat{f}_{\rm IIP, pred}$ & $\widehat{p}_{\rm IIP, pred}$ & $\widehat{e}_{\rm IIP, pred}$\\ 
\hline
$\mathcal{S}$ &6 & 1.6 & 0.55 & 0.829 & 0.221 & 0.944 & 0.261\\
&-- & 1.6 & 0.55 & 0.862 & 0.219 & 0.919 & 0.275\\
$\mathcal{S}_B$ &6 & 1.4 & 0.57 & 0.795 & 0.189 & 0.946 & 0.221\\
&-- & 1.8 & 0.52 & 0.85 & 0.174 & 0.889 & 0.235\\
$\mathcal{S}_{m,23.5}$ &4 & 1.1 & 0.56 & 0.821 & 0.272 & 0.913 & 0.355\\
&-- & 1.8 & 0.55 & 0.849 & 0.294 & 0.882 & 0.408\\
$\mathcal{S}_{m,24}$ &6 & 1.6 & 0.57 & 0.818 & 0.584 & 0.923 & 0.735\\
&-- & 1.6 & 0.57 & 0.787 & 0.488 & 0.891 & 0.674\\
$\mathcal{S}_{m,24.5}$ &4 & 1 & 0.55 & 0.755 & {\bf 0.612} & 0.904 & 0.807\\
&-- & 1.4 & 0.54 & 0.727 & 0.569 & 0.881 & 0.789\\
$\mathcal{S}_{m,25}$ &4 & 1.1 & 0.51 & 0.694 & 0.582 & 0.889 & 0.835\\
&-- & 1.2 & 0.51 & 0.6 & 0.543 & 0.856 & 0.845\\
$\mathcal{S}_{z,0.4}$ &6 & 1.4 & 0.59 & 0.715 & 0.245 & 0.724 & 0.552\\
&-- & 1.6 & 0.57 & 0.754 & 0.315 & 0.836 & 0.493\\
$\mathcal{S}_{z,0.6}$ &4 & 1 & 0.55 & 0.671 & 0.298 & 0.779 & 0.606\\
&-- & 1.4 & 0.57 & 0.658 & 0.375 & 0.846 & 0.67\\
\hline
\end{tabular}

$n_s=$ -- indicates that host-galaxy redshift is used as a covariate in the Random Forest classifier, and not to construct diffusion map.\\
$f^*_{\rm IIP}$ is computed on training set via 10-fold cross-validation.\\
$\widehat{f}_{\rm IIP, pred}$, $\widehat{p}_{\rm IIP, pred}$, and $\widehat{e}_{\rm IIP, pred}$ are evaluated on all data in the photometric set $\mathcal{P}$ not in the training set.
\end{table}

\section{Summary and Conclusions}
\label{sec:summary}

In this paper, we introduce the first use of semi-supervised classification for supernova typing.  Most of the previous methods have
relied on template fitting. Only recently, due in large part to the  Supernova Classification Challenge, other
statistics and machine learning methods been used for SN typing.  Our semi-supervised approach makes efficient use of 
the data
by using {\bf all} photometrically observed supernovae to find an appropriate representation for their classification.  
Also, we show that the complex variation in SN
light curves as a function of redshift is captured by this representation 
In this
manner, classification accuracy will improve as both the number of observed SNe grows and the parameters
such as redshift, stretch, and reddening are sampled more densely.  
It is not clear how this adaptation can occur for existing methods,
where either a fixed set of templates is used, or sets of summary statistics are extracted from each light curve independently.

Another advantage of our approach is the flexibility in the choice of distance measure, $s$, in the diffusion map construction.  In our analysis,
we used only the shapes of the light curves and colours of the SN to define $s$ (eq.~\ref{eqn:distband}) and also showed how this distance
can be modified if host-galaxy redshifts are available (eq. \ref{eqn:hostz}).  When using a diffusion map for SN classification, each 
astronomer is free to use their own choice of $s$, and presumably more sophisticated distance measures, such as ones
that include more 
context information or more accurately capture the inter-class differences in SN light curves, will perform better.

In applying our semi-supervised classification approach to data from the SN Classification Challenge, we find that results are
highly sensitive to the training set used.  We proposed a few spectroscopic follow-up strategies, discovering that deeper
magnitude-limited surveys obtain the best classification results--both for Type Ia and II-P SNe--despite 
accruing labels for a smaller number of supernovae.  Results show that our methods are competitive with the entrants of the SN Challenge 
as we obtain Type Ia purity/efficiency of 72\%/65\% on the photometric sample without using host redshifts, and 76\%/74\% using
host redshifts.  We hesitate to compare directly with results from that challenge due to large differences in the corrected set
of data studied in this paper.

Throughout this study, 
we attempted to avoid all use of SED templates.  However, there is some physical knowledge that 
is impervious to SED template problems, such as cosmological time dilation.
Indeed, our method of incorporating host-$z$ using
equation (\ref{eqn:hostz})  does account for time dilation: even if a high-$z$ type Ia SN
light curve appears like a low-$z$ type II light curve,  the degeneracy is broken because they are pulled apart in diffusion space
since their redshifts are different. Thus, our conclusion that deeper magnitude-limited surveys produce better 
training sets is not simply an artifact of neglecting to explicitly model time
dilation: deeper training sets still (tremendously) outperform 
shallower training sets even after incorporating host redshift (see Tables \ref{tab:zclassIa}--\ref{tab:zclassIIp}). The
improved performance is likely due both to training on lower S/N data
and capturing SED-dependent effects at high z.

Though our template-free method allows us to avert classification errors that arise through use of wrong or incomplete template
bases, as a trade-off we cannot extend results to redshifts outside of our
training data, and thus require deeper training sets.   This inability of our method to extrapolate beyond the training set is exhibited by the
poor performance  at high redshifts for the redshift-limited training sample (see Figures \ref{fig:fomz} and \ref{fig:fomz4}).
This suggests that our methods can be improved with increased use 
of physical modeling; indeed the future may lie in a combination of both semi-supervised learning and template methods.

In a future work, we plan to research such \emph{hybrid} models, which combine the semi-supervised approach presented in this paper
with physical SN models, in hopes to decrease the dependency of our results on the specific training set employed and to 
improve the overall classification accuracy.  One possible approach to this is to include, along with all of the observed SNe,
sets of supernovae simulated from different templates with different values of the relevant parameters.  This method would 
allow both the observed supernova light curves, along with the templates, to discover the optimal diffusion map
representation of the SNe.  Then, in the supervised part of the algorithm, the classification model would be able to learn
from both the spectroscopic data and the simulated supernovae, allowing the model to extend more widely (e.g., to higher 
redshifts than those sampled by the training set).


\section*{Acknowledgments}
J.W.R. acknowledges the generous support of a Cyber-Enabled Discovery and Innovation (CDI) grant (\#0941742) from the National Science Foundation.  Part of this work was performed in the CDI-sponsored Center for Time Domain Informatics (\url{http://cftd.info}).  P.E.F and C.M.S. acknowledge NSF grant 0707059 and NASA AISR grant NNX09AK59G.  D.P. is supported by an Einstein Fellowship. 

\appendix

\section{Classification Trees}
\label{ss:classificationTrees}

We start with a brief description of classification trees (\citealt{brei1984}).  Start with the $n$ supernovae with known class labels.  These data are pairs $(\mathbf{\Psi}_1,y_1),...,(\mathbf{\Psi}_{n},y_{n})$, where each $\mathbf{\Psi}_i$ is the $m$-dimensional diffusion map representation from equation (\ref{eqn:dmap2}), and each classification $y_i$ is coded to take on a value in $\{1,...,K\}$.  The classification tree consists of a series of recursive binary splits of the $m$-dimensional diffusion map space. Each split is performed on one coordinate, resulting in two nodes (two regions of input space).  The predicted class label $k \in \{1,...,K\}$, at each terminal node, $R_j$ is
\begin{equation}
\label{eqn:tree}
\widehat{h}_j = \arg\max_k \frac{1}{n_j} \sum_{\mathbf{\Psi}_i \in R_j} I(y_i = k)
\end{equation}
where $n_j$ is the number of training points in $R_j$ and $I(\cdot)$ is the indicator function.  Thus, the predicted class label at each node in the classification tree is the supernova type with the largest proportion of data points in $R_j$.  The tree classifier for any input $\mathbf{\Psi}$ is
\begin{equation}
\label{eqn:tree1}
\widehat{h}_{\rm tree}(\mathbf{\Psi},\mathbf{\Theta}) = \sum_j \widehat{f}_j I(\mathbf{\Psi} \in R_j),
\end{equation}
i.e., the prediction for $\mathbf{\Psi}$ is the predicted class for its terminal node, defined by equation (\ref{eqn:tree}).  In equation (\ref{eqn:tree1}), $\mathbf{\Theta}$ characterizes a tree in terms of the variable split and cutpoint at each node, and the predicted terminal node classifications.

To build the classification tree, at each successive node we choose the variable and splitting point that produces the largest decrease in misclassification rate, defined as
\begin{equation}
\label{eqn:misclass}
L(\mathbf{\Theta}) = \frac{1}{n}\sum_{i=1}^{n} I(y_i \ne \widehat{h}_{\rm tree}(\mathbf{\Psi}_i))
\end{equation}
Other versions of classification trees use Gini index (used by CART) or entropy (used by C4.5) to determine the splitting points.  The splitting process is repeated until $\min_j n_j \le n_{\min}$, giving us the estimated model (\ref{eqn:tree1}).

Tree models are simple, yet powerful, nonparametric classifiers.  They work well even when the true function $h$ that relates the input space to the class labeling is complicated, generally yielding estimates with very small bias.  Furthermore, there are fast, reliable algorithms for fitting trees and prediction is very simple.  However, tree models tend to have high variance.  Small changes in the SN training set, $\mathbf{\Psi}_1,...,\mathbf{\Psi}_n$, can produce very different estimated tree structure.  This is a result of the hierarchical nature of the tree model: small differences in the top few nodes of a tree can produce wildly different structure as those perturbations are propagated down the tree.  To reduce this instability, we use the random forest approach, which we now describe.

\section{Random Forests}
\label{ss:randomForests}
Random forest is a committee method that builds a large collection,
$B$, of decorrelated classification trees, each one fit on a subset
of the data, and then averages them to obtain a final predictive
model, $\widehat{h}_{\rm rf}$ with reduced variance over any single
classification tree.  Random forest is an improvement to bagging
(bootstrap aggregation, \citealt{brei1996}), a method that averages
the predictions of $B$ trees fit to bootstrapped samples of the data,
because it attempts to decorrelate the $B$ trees (without increasing
the variance too much) by selecting a random set $r<m$ of the input
coordinates as candidates for splitting at each node during the tree
building process.  The net result of building decorrelated
classification trees is that the final, averaged model will have
smaller variance than the bagging model (see \citealt{hast2009},
ch. 15 for a discussion). The observations and coordinates that are
not included in a given tree are referred to as ``out of bag.''

After growing $B$ decorrelated classification trees, $\widehat{h}_{\rm tree}(\mathbf{\Psi},\mathbf{\Theta}_1),...,\widehat{h}_{\rm tree}(\mathbf{\Psi},\mathbf{\Theta}_B)$, we let them vote on the class label to determine the classification of supernova $\mathbf{\Psi}$.  The predicted classification for $\mathbf{\Psi}$ is the majority vote of these trees,
\begin{equation}
\label{eqn:rf}
\widehat{h}_{\rm rf}(\mathbf{\Psi}) = {\rm majority \hspace{.04in} vote} \left\{ \widehat{h}_{\rm tree}(\mathbf{\Psi},\mathbf{\Theta}_b)\right\}_{b=1}^B
\end{equation}
where ties are broken at random.  Alternatively, we can keep track of the number of `votes' for each class to estimate a measure of classification probability of each class for a supernova.  To fit random forest models to the diffusion map representations of SNe, we use the {\tt randomForest R} package\footnote{The {\tt randomForest R} package is available at \url{http://cran.r-project.org/web/packages/randomForest/}.}.  An advantage to random forest is the relative robustness of the estimates to choices of the tuning parameters compared to other nonparametric classification techniques. In practice we build $B=500$ trees, select $r = \sqrt{m}$ coordinates as splitting candidates at each node, set the minimum node size to $n_{\min} =1$ and take bootstrap samples of size $n$ (where $n$ is the number of observed supernovae).  The dimensionality, $m$, of the diffusion map space and the diffusion map tuning parameter $\epsilon$ can both affect classification results, so we optimize our training set misclassification error over $(\epsilon,m)$.

\bibliography{SNclass}

\label{lastpage}

\end{document}